\documentclass[12pt]{article}

\usepackage{amsmath,epsfig,amssymb,amsfonts,amsthm,epsfig,verbatim, enumitem}
\usepackage{color,graphicx,lscape,longtable,mathabx,natbib}
\usepackage{multirow}
\usepackage{booktabs,siunitx}
\usepackage{float}
\usepackage{caption}
\usepackage{subfigure}
\usepackage{threeparttable}

\DeclareMathOperator*{\argmin}{arg\,min}

\usepackage{longtable}

\newtheorem{Theorem}{Theorem} %[section]
\newtheorem{Ass}{Assumption} %[section]
\newtheorem{Lemma}[Theorem]{Lemma}

\newtheorem{Th}{\underline{\bf Theorem}}

\newtheorem{Rem}{\underline{\bf Remark}}

\newtheorem{Cor}{Corollary} %[section]

\def\bse{\begin{eqnarray*}}
\def\ese{\end{eqnarray*}}
\def\be{\begin{eqnarray}}
\def\ee{\end{eqnarray}}
\def\bsq{\begin{equation*}}
\def\esq{\end{equation*}}
\def\bq{\begin{equation}}
\def\eq{\end{equation}}

\def\var{\hbox{var}}

\def\wh{\widehat}
\def\wt{\widetilde}

\def\n{\nonumber}

\def\sumi{\sum_{i=1}^n}

\def\trans{^{\rm T}}

\def\bomega{\boldsymbol\omega}

\def\bnu{\boldsymbol\nu}
\def\beps{\boldsymbol\epsilon}

\def\bb{{\boldsymbol\beta}}

\def\bg{{\boldsymbol\gamma}}

\def\0{{\bf 0}}

\def\U{{\bf U}}

\def\v{{\bf v}}

\def\D{{\bf D}}

\def\I{{\bf I}}
\def\M{\mbox{ $\mathcal{M}$}}
\def\M{{\bf M}}

\def\K{{\bf K}}

\def\brho{{\boldsymbol\rho}}

\def\U{{\bf U}}
\def\S{{\bf S}}

\def\v{{\bf v}}
\def\W{{\bf W}}
\def\Q{{\bf Q}}

\def\X{{\bf X}}

\def\S{{\bf S}}

\def\I{{\bf I}}

\def\Y{{\bf Y}}

\def\y{{\bf y}}
\def\Z{{\bf Z}}

\def\bSig{{\bf \Sigma}}

\def\bq{\begin{equation}}
\def\eq{\end{equation}}

\def\trans{^{\rm T}}

\def\log{{\rm log}}

\def\squarebox#1{\hbox to #1{\hfill\vbox to #1{\vfill}}}
\def\btheta{{\boldsymbol \theta}}

\def\var{\hbox{var}}

\def\bse{\begin{eqnarray*}}
\def\ese{\end{eqnarray*}}
\def\be{\begin{eqnarray}}
\def\ee{\end{eqnarray}}
\def\bsq{\begin{equation*}}
\def\esq{\end{equation*}}
\def\bq{\begin{equation}}
\def\eq{\end{equation}}

\def\trans{^{\rm T}}

\def\boxit#1{\vbox{\hrule\hbox{\vrule\kern6pt\vbox{\kern6pt#1\kern6pt}\kern6pt\vrule}\hrule}}

\usepackage{xr}
\usepackage{prettyref}
\begin{document}
\thispagestyle{empty}
\allowdisplaybreaks
\renewcommand {\thepage}{}
\include{titre}
\pagenumbering{arabic}
\begin{center}
{\Large{\bf Inference in High-Dimensional Linear Measurement Error
    Models}}
\vskip 1mm {\bf Mengyan Li, Runze Li and Yanyuan Ma}

Department of Statistics, The Pennsylvania State University, University Park,
PA 16802
\end{center}
\baselineskip 23pt
\vskip 1mm

\begin{abstract}
For a high-dimensional linear model with a finite number of covariates measured
with error, we study statistical inference on the parameters
associated with the error-prone covariates, and propose a new corrected
decorrelated score test and the corresponding one-step estimator. We further
established asymptotic properties of the newly proposed test statistic and the
one-step estimator. Under local alternatives, we show that the limiting
distribution of our corrected decorrelated score test statistic is non-central
normal. The finite-sample performance of the proposed inference procedure is
examined through simulation studies. We further illustrate the proposed
procedure via an empirical analysis of a real data example.


\end{abstract}

\section{Introduction}
High dimensional data becomes more and more common in diverse fields such as
computational biology, economics and climate science. Many statistical
procedures have been developed for analysis of high dimensional data.
%
%Extensive research has been devoted to developing variable selection techniques
%and studying their statistical properties for high dimensional regression
%problems. For example, Lasso \citep{tibshirani1996regression}, SCAD
%\citep{fan2001variable}, elastic net \citep{zou2005regularization} and Dantzig
%selector \citep{candes2007dantzig} have been studied carefully in the
%literature and are widely used.
However, most of them often assume that all covariates are measured
accurately. In reality, measurement errors are ubiquitous in many
high-dimensional problems, for example, measurements of
  gene expression with cDNA or oligonucleotide arrays
  \citep{rocke2001model} and sensor network data
  \citep{slijepcevic2002location}. 
This work was motivated by an empirical analysis of a real data set in
Section \ref{subsec:realdata}, where both finite-dimensional
phenotypic covariates and high-dimensional SNPs are available and one
of the phenotypic covariates is of clinical interest but measured with
error.

The classical measurement error models, where the number of covariates
$p$ is fixed or is smaller than
 the sample size $n$, have been studied systematically, see
 \cite{fuller2009measurement},
 %{\color{blue}\cite{nakamura1990corrected}}, 
 \cite{carroll2006measurement}, 
 \cite{grace2016statistical} and \cite{mali2010}.
Penalized methods have been developed for high-dimensional linear measurement
error models with $p>n$. Consider the model
\be \label{eq:model_all}
\Y = \X\btheta_0 + \beps, \ \ \text{and}  \ \ \W = \X + \U,
\ee
where random vectors $\Y, \beps \in \mathbb{R}^n$, the $n\times p$
matrix $\X$ is unobservable, $\W$ 
is its observed surrogate, and the matrix $\U$ is random noise, i.e.
measurement error. 
 This is a difficult problem. In fact, 
even in the absence of measurement error,
\cite{zhao2006model} and \cite{meinshausen2006high} showed that the
Lasso or Dantzig selector often fails in identifying significant
covariates in high-dimensional models. 
With measurement error, \cite{rosenbaum2010sparse} showed that  the
true selection is likely to be outside of the feasible set of the
Dantzig selector. \cite{sorensen2015measurement} analyzed the impact
of measurement error on the standard Lasso and showed that
treating $\W$ as the true $\X$  leads to erroneous results.

To correct the bias caused by  the measurement error $\U$,  a
corrected objective function 
%{\color{blue} first proposed by \cite{nakamura1990corrected}} 
is
\bse
\frac{1}{2} \btheta\trans \wh\bSig \btheta - \frac{1}{n}\W\trans\y + P_{\lambda}(\btheta),
\ese
where $P_{\lambda}(\btheta)$ is a penalty with tuning parameter
$\lambda$, $\wh\bSig = \W\trans\W/n - \D$, and $\D$ is the $p\times p$
covariance matrix of $\U_i$.
Since $\wh\bSig$ can have negative eigenvalues
when $p$ is larger than $n$,
the loss function $ \btheta\trans \wh\bSig \btheta/2 - \X\trans\y/n $ is no longer convex.
To overcome the difficulties caused by the non-convexity,
\cite{loh2012} proposed a projected gradient
descent algorithm that finds a possible local optimum with strong performance guarantees. 
\cite{chen2013noisy} developed a simple variant of orthogonal
matching pursuit algorithm that performs at the minimax optimal
rate.
Later,
\cite{belloni2017linear} proposed the compensated matrix uncertainty (MU)
selector, which can be written as a second-order cone programming
minimization problem and the estimator attains the minimax efficiency
bound.
 \cite{loh2017support}
developed a primal-dual witness proof framework to establish the
estimator error bounds in different norms
in general sparse regression problems with non-convex loss function and penalty.
This work does not require the typical incoherence
condition, but need to impose the constraint $\|\btheta_0\|_1<R$.
\cite{datta2017cocolasso} proposed CoCoLasso estimator which forces
the non-convex problem to be convex by applying a nearest positive
semi-definite matrix projection operator to $\wh\bSig$, which can be
solved by the ADMM algorithm, and analyzed its error bounds with
deterministic design matrix $\X$.
Under a slightly stronger sparsity conditions, the asymptotic
sign-consistency properties were established.

The aforementioned works focus on the theory and numerical algorithms of regularization methods rather than statistical inference. It is important to quantify the uncertainty of an estimator in high
dimensional linear measurement error models.
Recently, significant progress has been made regarding
hypothesis testing on low dimensional sub-parameters
in high dimensional sparse models.
From a semiparametric perspective,
the challenges in these problems lie in how to handle the effect of
high-dimensional nuisance parameters and correct the bias of the
estimators for the low dimensional parameters of interest caused by the
penalty.
\cite{zhang2014confidence} proposed a low dimensional projection (LDP)
approach to construct bias-corrected linear Lasso estimator and
corresponding confidence intervals without assuming the uniform signal
strength condition \citep{wainwright2009information}.
\cite{vandegeer2014} exploited the idea of inverting the
Karush-Kuhn-Tucker characterization to desparsify Lasso, which
essentially leads to the same results as in \cite{zhang2014confidence} for a linear model.
\cite{javanmard2014confidence} proposed to debias the Lasso estimator
by adding a term proportional to the subgradient of the $\ell_1$ norm at the Lasso
solution, and the confidence intervals constructed based on the debiased
estimator have nearly optimal size. All these works assume either linear or generalized linear models.
\cite{ning2017general} provided a general framework for
high-dimensional inference by proposing a decorrelated score function.
By applying a decorrelation operation on the high-dimensional score functions,
the derived decorrelated score function is uncorrelated with the nuisance score function.
In this case,  the efficiency of the estimators for the parameters of interest will not be impaired
provided that the estimators for the nuisance parameters are consistent at sufficient rate.
%Their method is generally applicable to a wide variety of high dimensional sparse models with different loss functions and convex or non-convex penalties.

Inference for high dimensional measurement error models is believed to be a difficult topic due to the bias
and lack of power introduced by measurement error as well as high dimensional nuisance parameters. 
Recently, \cite{belloni2017inference} constructed simultaneous confidence regions
for the parameters of interest in high-dimensional linear models with error-in-variables using multiplier bootstrap.
\cite{wang2019rate}  employed a de-biasing approach and constructed component-wise confidence intervals in a sparse high-dimensional linear regression model when some covariates of the design matrix are missing completely at random.
%, which can be viewed as a certain kind of measurement errors. }
In this paper, we consider the
setting where only a fixed number of covariates are measured with error and our goal
is to develop statistical inference procedures for the coefficients of these covariates.
In practice, it is common that not all covariates are corrupted.
For example, in the real data example analyzed in Section 4.2, covariates
such as gender and age are measured precisely.  Moreover, it
is in general very difficult to find a good
estimate for the $p \times p $ covariance matrix $\D$ of measurement
error without any strong and restrictive assumptions.

We extend the inference results of low dimensional  linear measurement
  error models %in \cite{nakamura1990corrected} 
to high dimensional
  settings, which is important yet challenging, and requires vastly
  different treatments.
In the spirit of semiparametrics, we employ decorrelation operation to
control the impact of high-dimensional nuisance parameters, and
construct a corrected decorrelated score function for the parameters of
interest.
The performance of the corrected decorrelated score test relies on the
convergence rate of the initial estimator. The asymptotic normality of
the corrected decorrelated score test statistic holds provided that the initial
estimator is statistically consistent at certain rate. Here, we take the CoCoLasso
estimator \citep{datta2017cocolasso} as an example. Indeed,
any estimator with sufficient convergence rate can be served as the initial
estimator in forming the decorrelated score function.
Different from the settings in \cite{datta2017cocolasso}, we assume that
the design is random and sub-Gaussian, and only a fixed number of
covariates, without loss of generality, one covariate,  is
measured with error.
We rederive the theoretical properties of  the CoCoLasso estimator in our new settings, which is one of the contributions of this work. 
Our corrected decorrelated score test statistics retain power under the local
alternatives around $0$, because we essentially do not impose any penalty on the parameter of interest in the construction.
%We  modify the theoretical proofs of \cite{datta2017cocolasso}, and
%hence the error bounds are also different in terms of certain constants.
We further construct confidence intervals by proving the limiting
distribution of the one-step estimator, which is semiparametrically efficient. 
%, which is the root of the estimating equation based on the estimated decorrelated score function.
Note that although we write our development for one variable with
measurement error, the proposed method is directly applicable to
a finite number of covariates with measurement error naturally.

Our work extends the key idea of semiparametrics to inference in high dimensional
linear measurement error models.
We handle the sparsity assumptions differently from \cite{belloni2017linear} and \cite{loh2012}, and extend the results in \cite{datta2017cocolasso} to random sub-Gaussian designs. 
Although a general framework of inference was provided in 
\cite{ning2017general}, the existence of measurement errors
imposes many special challenges in methodology and theoretical proofs, which requires innovative technical treatments, as illustrated in the main text of the paper.	
Compared to \cite{belloni2017inference},
we avoid solving estimating equations completely.
Our one-step estimator has the same limiting distribution  as that of the root of estimating equations but is much easier to compute.

We specify the model for high-dimensional data with one
covariate with measurement error and develop the methodology in Section \ref{sec:2}, which includes construction of the corrected
decorrelated score function, statistical properties of the initial
estimator as well as the algorithm.
Technical conditions, asymptotic properties of the score test
statistic and the one-step estimator are established in Section
\ref{sec:3}.
To assess the performance of our method, we conduct simulation studies
and perform an empirical data analysis in Section \ref{sec:4}.

\label{subsec:notation}
{\bf Notations and Preliminaries}: Before we pursue further, let us introduce some notation and some
preliminaries. For a vector $\v = (v_1, \dots, v_p)\trans \in \mathbb{R}^p$, we
define $\|\v\|_0 = |\rm{supp}(\v)|$, where $\rm{supp}(\v) = \{j: v_j
\neq 0\}$
and $|A|$ is the cardinality of a set $A$.
Denote $\|\v\|_{\infty} = \max_{1\leq j \leq p}|v_j|$ and $\v^{\otimes2} = \v\v\trans$.
For $S \subseteq \{1, \dots, p\}$, let $\v_S = \{v_j: j \in S\}$ and $S^C$ be the complement of $S$ .
For a matrix $\M = [\M_{jk}]$, let $\|\M\|_{\max} = \max_{j,
  k}|M_{jk}|$, $\|M\|_{\infty} =\max_j \sum_{k}|M_{jk}|$ and
$\M^{\otimes2} = \M\M\trans$.
If $\M$ is symmetric, then $\lambda_{\min}(\M)$ and
$\lambda_{\max}(\M)$ are the minimal and maximal eigenvalues of $\M$.
For two positive sequences $a_n$ and $b_n$, we use $a_n \precsim b_n$
to denote $a_n \leq Cb_n$ for some constant $C>0$,
and use $a_n \asymp b_n$ to denote $C\leq a_n/b_n \leq C^{\prime}$ for
some constants $C, C^{\prime}>0$.
Denote $\Phi(\cdot)$ to be  the cumulative distribution function of
the standard normal distribution.
For simplicity, we use $E(\cdot)$ and $\Pr(\cdot)$ to denote the
expectation and probability calculated under the true model,
respectively.

The sub-exponential norm of a random variable $X$ is defined as $\|X\|_{\psi_1} =
\sup_{q\geq1} q^{-1}\{E(|X|^q)\}^{1/q}$.
Note that $\|X\|_{\psi_1} < C_1$ for some constant $C_1$, if $X$ is sub-exponential.
The sub-Gaussian norm of $X$ is defined as $\|X\|_{\psi_2} =
\sup_{q\geq1} q^{-1/2}\{E(|X|^q)\}^{1/q}$.
Note that $\|X\|_{\psi_2} < C_2$ for some constant $C_2$, if $X$ is sub-Gaussian.
More properties regarding sub-exponential and sub-Gaussian random
variables are given in Appendix G.1 in the supplementary materials.

\section{Model Setup and Proposed Method}
\label{sec:2}

\subsection{Model Specification}
\label{subsec:model}

Suppose that $\{Y_i, W_i, \Z_i\}$, $i=1,\dots,n$, is an independent and identically distributed sample
from a linear model with one of the covariates measured with additive error
\be\label{eq:model}
Y_i=\beta_0 X_i + \bg_0\trans\Z_i + \epsilon_i \ \ \text{and} \ \ W_i=X_i+U_i.
\ee
Covariate $X_i \in \mathbb{R}$ is unobservable, and $W_i$ is its error-prone surrogate.
Covariate vector $\Z_i \in \mathbb{R}^{p-1}$ is measured precisely.
Assume that $(X_i, \Z_i\trans)\trans$ is sub-Gaussian element-wise with mean $\0$ and unit diagonal covariance matrix.
To exclude the intercept term in the model, we let the response $Y_i$ have mean 0 as well.
The regression error $\epsilon_i$ is sub-Gaussian with
mean 0, variance $\sigma_{\epsilon}^2$, and sub-Gaussian norm $K_{\epsilon}$.
The measurement error $U_{i}$ is also
sub-Gaussian with mean 0, variance $\sigma_U^2$, and sub-Gaussian norm $K_U$.
It is independent of $\epsilon_i$, $X_i$ and $\Z_i$.
As in the literature, we assume that $\sigma_U^2$ and $E(U_i^4)$ are known.

Let $\Y=(Y_1, \dots, Y_n)\trans$, $\X=(X_1, \dots, X_n)\trans$, $\W=(W_1, \dots,
W_n)\trans$ and $\Z=(\Z_1, \dots, \Z_n)\trans$ denote the corresponding vector or
matrix version of $n$ samples.
In practice, we only need to center all variables, and standardize the columns of the data matrix such that $\sum_{i=1}^{n}Z_{ij}^2/n =1$ and $\sum_{i=1}^{n}W_{i}^2/n =1+\sigma_U^2$ for $j=1, \dots, p-1$ and $i=1,\dots, n$ .

For the purpose of theoretical proofs, we have the following standard assumptions.
\begin{Ass}
\label{ass:model}
Assume that
\begin{itemize}
    \item[{\rm (i)}] $2\kappa \leq \lambda_{\min} [E\{(X_i, \Z_i\trans)^{\trans\otimes2}\}]
    \leq \lambda_{\max} [E\{(X_i, \Z_i\trans)^{\trans\otimes 2}\}]\leq 2/\kappa$
    for some constant $\kappa >0$;
    \item[{\rm (ii)}]$\|Z_{ij}\|_{\psi_2}$ and $\|X_i\|_{\psi_2}$ are uniformly bounded by some constant $K$ for $j=1, \dots, p-1$;
    \item[{\rm (iii)}] The true parameter $\btheta_0 = (\beta_0,
      \bg_0\trans)\trans$ is sparse with
      support $S$, and $|S| = s_0$;
      Let $\|\btheta_0\|_\infty \leq K_0$, where $K_0$ is a positive constant;
    \item[{\rm (iv)}]
     $E(X_i\Z_i\trans)\{E(\Z_i^{\otimes2})\}^{-1}$
     is sparse with support $S^{\prime}$ and $|S^{\prime}| = s^{\prime}$.
     Moreover, \par
    $\|E(X_i\Z_i\trans)\{E(\Z_i^{\otimes2})\}^{-1}\|_1 \leq
    K_{\bomega}$ for some constant $K_{\bomega} > 0$.

\end{itemize}
\end{Ass}

In Assumption \ref{ass:model},
${\rm (i)}$ and ${\rm (ii)}$ are common assumptions for high
dimensional random designs.
Assumption ${\rm (iii)}$ is about the sparsity of the true model
(\ref{eq:model}). Instead of assuming $\|\btheta_0\|_1$ is bounded, we
only assume the $l_{\infty}$ norm of $\btheta_0$ is bounded.
Assumption ${\rm (iv)}$ is crucial in the inference framework of
\cite{ning2017general}.
When conducting decorrelation operation, their key assumption is that
the projection of the score function for $\beta$ to the linear space
spanned by the nuisance score functions for $\bg$, denoted as
$\Lambda_{\bg}$, is identical to the projection of the score function
for $\beta$ to a low dimensional subspace of $\Lambda_{\bg}$.
More details about the motivation of sparse projection and the
formation of $E(X_i\Z_i\trans)\{E(\Z_i^{\otimes2})\}^{-1}$
will be discussed in Section \ref{subsec:dsf}.

%Under the above assumptions, we know $W_i$ is also sub-Gaussian and $\|W_i\|_{\psi_2}$ is bounded by $K+K_U$.
Our goal is to test the hypothesis $H_0: \beta_0=\beta^*$ and
construct valid confidence intervals for $\beta_0$ when the dimension
of $\btheta_0=(\beta_0, \bg_0\trans)\trans$ is much larger than the
sample size $n$, that is, $p \gg n$.
Note that when $\beta^* = 0$, under the null hypothesis, the model
degenerates to a linear model without measurement error, hence testing
procedures for high dimensional sparse linear models can be applied. In
this paper, we consider a general hypothesis test setting where
$\beta^* \in \mathbb{R}$.

%\section{Methodology}
%\label{sec:3}
\subsection{Corrected Decorrelated Score Function}
\label{subsec:dsf}
If covariate $X$ is observed with no measurement error,
it is known that the loss function based on least squares is  $\btheta\trans\bSig\btheta/2-\brho\trans\btheta$,
where $\bSig = (\X, \Z)\trans(\X, \Z)/n$ and $\brho = (\X, \Z)\trans \Y/n$.
For our corrupted data $(\Y, \W, \Z)$,
as emphasized above, instead of treating $\W$ as $\X$ in the loss function directly, we define the corrected loss function as
\be
\label{eq:correctedloss}
l(\btheta)
= \frac{1}{2}\btheta\trans\wh\bSig\btheta-\wh\brho\trans\btheta,
\ee
where
\bse
\wh\bSig= \frac{1}{n}(\W, \Z)\trans(\W, \Z)-
\begin{pmatrix}
\sigma_U^2 & \0 \\
\0 & \0
\end{pmatrix} \ \ \text{and} \ \ \wh\brho &=& \frac{1}{n} (\W, \Z)\trans\Y.
\ese
%and $l_i(\btheta)= (W_i^2-\sigma_U^2)\beta^2/2+W_i\Z_i\trans\bg\beta +(\Z_i\trans\bg)^2/2-W_iY_i\beta-Y_i\Z_i\trans\bg$.
By assumption, $U_i$ is independent of $X_i$, $\Z_i$ and $\epsilon_i$, it is easy to verify that
$E(\wh\bSig)=E(\bSig)$ and $E(\wh\brho) = E(\brho)$.

The gradient of the loss function plays an important role in statistical analysis.
Because our corrected loss function is no longer the log-likelihood,
we name it the gradient corrected score function, which
has the form $\S_{\btheta}(\btheta)=n^{-1}
\sumi\S_{i\btheta}(\btheta)= \wh\bSig\btheta-\wh\brho$.
Because we aim at conducting inference on the parameter $\beta$, we
treat the $p-1$ dimensional parameter $\bg$ as nuisance.
Then the corrected score function can be decomposed as
\bse
\S_{\btheta}(\btheta) =
\begin{pmatrix}
S_{\beta}(\beta, \bg)\\
\S_{\bg}(\beta, \bg)
\end{pmatrix}
=
\begin{pmatrix}
\wh\Sigma_{11} \beta+ \wh\bSig_{12}\bg - \wh\rho_1 \\
\wh\bSig_{21} \beta+ \wh\bSig_{22}\bg  - \wh\brho_2
\end{pmatrix},
\ese
where  $\wh\Sigma_{11}=\W\trans\W/n - \sigma^2_U$,
$\wh\bSig_{12}=\W\trans\Z/n$, $\wh\bSig_{21}=\Z\trans\W/n$,
$\wh\bSig_{22}=\Z\trans\Z/n$, $\wh\rho_1 = \W\trans\Y/n$ and
$\wh\brho_2 = \Z\trans\Y/n$.

Similar to the standard score function,
it can be easily verified that $E\{\S_{i\btheta}(\btheta_0)\}=\0$.
Define the $p\times p$ corrected score covariance matrix as
\bse
\I(\btheta) =E\{\S_{i\btheta}(\btheta)\S_{i\btheta}(\btheta)\trans\} =
\begin{pmatrix}
I_{\beta\beta} & \I_{\beta\bg} \\
\I_{\bg\beta} & \I_{\bg\bg}
\end{pmatrix}.
\ese
Note that the covariance matrix $\I(\btheta)$ is no longer equal to
$E\{\partial\S_{i\btheta}(\btheta)/\partial\btheta\trans\}$ due to the
bias correction procedure in constructing the loss function.
In fact, the matrix $\I(\btheta)$ has more complex form.
With standardized data matrix $(\X, \Z)$, by simple calculations we obtain that
 \be
 \label{eq:matrixI}
\I(\btheta)
 = \begin{pmatrix}
 (\sigma_{\epsilon}^2  + \beta^2\sigma_U^2) +
 \sigma_{\epsilon}^2\sigma_U^2 +\beta^2E(U_i^4) - \beta^2\sigma_U^4 &
 (\sigma_{\epsilon}^2  + \beta^2\sigma_U^2)E(X_i\Z_i)\trans  \\
 (\sigma_{\epsilon}^2  + \beta^2\sigma_U^2)E(X_i\Z_i)  &
 (\sigma_{\epsilon}^2  + \beta^2\sigma_U^2)E(\Z_i\Z_i\trans)
 \end{pmatrix}.
\ee
To control the impact of high-dimensional nuisance parameter $\bg$ on the inference of the parameter of interest $\beta$,
we define the corrected decorrelated score function for $\beta$ as
\bse
S(\beta, \bg) = S_{\beta}(\beta, \bg) - \bomega\trans
\S_{\bg}(\beta, \bg),
\ese
where $\bomega\trans=\I_{\beta\bg}\I_{\bg\bg}^{-1} = E(X_i\Z_i\trans)E(\Z_i\Z_i\trans)^{-1} = E(W_i\Z_i\trans)E(\Z_i\Z_i\trans)^{-1}$.
Under the assumption that the minimal eigenvalue of $E\{(X_i, \Z_i\trans)^{\trans\otimes{2}}\}$ is bounded and bounded away from $0$,
it is easy to show that the $(p-1) \times (p-1)$ matrix $E(\Z_i\Z_i\trans)$ is invertible.
Note that this construction ensures that $S(\beta, \bg)$ is uncorrelated with the nuisance score function $\S_{\bg}(\beta, \bg)$,
i.e. $E\{S(\beta_0, \bg_0)\S_{\bg}(\beta_0, \bg_0)\}=\0$.
The detailed verification is in Appendix D.1 in supplementary materials.
We denote the variance of $S(\beta, \bg)$ as $\sigma_{\beta\mid \bg}^2$, and it is easy to show that
\be\label{eq:sigmaS}
\sigma^2_{\beta|\bg} = I_{\beta\beta} -
\I_{\beta\bg}\I_{\bg\bg}^{-1}\I_{\bg\beta}.
\ee

Under the null hypothesis $H_0: \beta_0=\beta^*$, to construct score
test statistic, we need to find estimators for the nuisance parameter
$\bg$ and the $p-1$ dimensional vector $\bomega$.
For $\bg$,  we can use any consistent
estimator $\wt\bg$ with sufficient convergence rate due
  to the decorrelation operation.
More details about $\wt\bg$ as well as the initial estimator
$\wt\beta$ for $\beta$ will be discussed in Section \ref{sec:ini}.
For $\bomega$, an intuitive estimator is its sample version $\wh\bSig_{12}\wh\bSig_{22}^{-1}$.
However, matrix $\wh\bSig_{22}$ is not invertible when $p-1 > n$.
\cite{ning2017general} imposed sparsity assumption on $\bomega$ to control the estimation error.
Many different penalized methods can be applied
to obtain a sparse estimator of $\bomega$.
For example, the Dantzig type estimator $\wh\bomega$ can be obtained as follows:
\be
\label{eq:dantzig}
\wh\bomega=\argmin \|\bomega\|_1 \ \ \text{s.t.}  \ \ \|\wh\bSig_{12}
- \bomega\trans\wh\bSig_{22}\|_{\infty}\leq \lambda^{\prime},
\ee
where $\lambda^{\prime}$ is a tuning parameter.
Note that in our model $\wh\bSig_{12}$ and $\wh\bSig_{22}$ do not depend on $\btheta$.
Then the estimated corrected decorrelated score function is
defined as
$\wh S(\beta, \wt\bg) = S_{\beta}(\beta, \wt\bg) - \wh\bomega\trans \S_{\bg}(\beta, \wt\bg)$.

Under the null hypothesis,  we construct the test statistic as
$\wh T_n = n^{1/2}\wh S(\beta^*, \wt\bg)(\wh\sigma^2_{\beta|\bg, H_0})^{-1/2}$,
where
\be
\label{eq:sigmacond}
\wh \sigma^2_{\beta|\bg, H_0}
&=&\{\wh I_{\beta\beta} - \wh\bomega\trans\wh \I_{\bg\beta} \n\}|_{\beta=\beta^*} \\
&=&(\wh\sigma_{\epsilon, H_0}^2 + \beta^{*2}\sigma_U^2)(1 -
\wh\bomega\trans\wh\bSig_{21}) + \beta^{*2}E(U_i^4) +
\wh\sigma_{\epsilon, H_0}^2\sigma_U^2 - \beta^{*2} \sigma_U^4,
\ee
and
$\wh\sigma_{\epsilon, H_0}^2 = n^{-1} \sum_{i=1}^n(Y_i - \beta^* W_i - \wt\bg\trans\Z_i)^2 - \beta^{*2}\sigma_U^2$.
The detailed derivation is given in Appendix D.2 in supplementary materials.
Under some assumptions we will specify in Section \ref{sec:4},
the test statistic $\wh T_n$ is asymptotically standard normal, see Corollary \ref{cor:scoretest}.

For confidence interval construction,
define the one-step estimator for $\beta$ as the root of the first
order approximation of the approximately unbiased estimating equation
$\wh S(\beta, \wt\bg) = 0$ around the initial estimator $\wt\beta$, i.e.,
\bse
\wh\beta &=& \wt\beta - \wh
S(\wt\btheta)/\{\partial \wh S(\beta, \wt\bg)/\partial \beta\}|_{\beta=\wt\beta} \\
&=& \wt\beta - \wh S(\wt\btheta)/ (\wh\bSig_{11}- \wh\bomega\trans\wh\bSig_{21}).
\ese
Of course,  we could use the true root of $\wh S(\beta, \wt\bg) = 0$ as
$\wh \bb$. Here, we choose to use the one-step update 
for its computational simplicity. In fact, 
we have proved that the asymptotic distribution 
of the one-step estimator is identical to
that of the true root because we have a relatively good initial estimator
$\wt \bb$.
%As mentioned in \cite{ning2017general}, the reason why not to
%solve $\wh S(\beta, \wt\bg) = 0$ directly is because it may have multiple roots.
We will show that the one-step estimator
$\wh\beta$ is consistent and asymptotically normal with
asymptotic variance $\sigma_\beta^2$
under suitable assumptions in Theorem \ref{th:asynormal2}.
Hence, the $(1-\alpha)100\%$ confidence interval for $\beta_0$ can be
constructed as $\left(\wh\beta - z_{\alpha}\sqrt{\wh\sigma_\beta^2/n},
\wh\beta + z_{\alpha}\sqrt{\wh\sigma_\beta^2/n}  \right)$,
where $\Phi(z_{\alpha}) = 1-\alpha/2$, and $\wh\sigma_\beta^2$ is an
estimate of $\sigma_\beta^2$ whose specific form is given in Theorem
\ref{th:asynormal2}.

\subsection{Initial Estimator}\label{sec:ini}

In the literature, estimation theories under different assumptions have
been developed for model (\ref{eq:model_all}), where all covariates
are measured with error, see \cite{loh2012}, \cite{chen2013noisy},
\cite{belloni2017linear}, \cite{loh2017support} and
\cite{datta2017cocolasso}.
With slight modifications, these methods can all be applied to our
model to construct desired initial estimators.
Here, we take CoCoLasso estimator proposed by
\cite{datta2017cocolasso} as an example to show how the convergence
performance of the initial estimator affects the inferential results of $\beta$.

The CoCoLasso estimator is defined as
\be
\label{eqn:CCL}
\wt\btheta = \argmin_{\btheta} \frac{1}{2}
\btheta\trans\wt\bSig\btheta - \wh\brho\trans \btheta + \lambda
\|\btheta\|_1,
\ee
where $\wt\bSig = (\wh\bSig)_+$ and $\lambda$ is a tuning parameter.
The nearest positive semi-definite matrix projection operator
$(\cdot)_+$ is defined as follows: for any matrix $\K$,
\bse
(\K)_+=\argmin_{\K_1 \geq \0} \|\K-\K_1\|_{\rm{max}}.
\ese
The ADMM algorithm is used to find the nearest positive semi-definite
matrix. For more details, see \cite{fan2016multitask} and
\cite{datta2017cocolasso}.

As mentioned in the introduction, since we consider sub-Gaussian design with fixed number of covariates measured with error, which is different from the settings in \cite{datta2017cocolasso},
we modified their theoretical proofs under our settings and the error bounds are different in terms of certain constants.
We give the $l_1$, $l_2$ and prediction error bounds of $\wt\btheta$
in the following Lemma.
\begin{Lemma}
\label{lemma:theta}
Let $\lambda = C_{\lambda}s_0\sqrt{n^{-1}\log p}=o(1)$.
For $C_{\lambda} > \max\left( 8K_0K_2/C^{\prime\prime}, 8\sqrt{2}K_0K_3/\sqrt{C^{\prime\prime}} \right)$ and
$\lambda \leq \min(8K_1, 16KK_{\epsilon}, 8K_0K_2, 8K_0K_3)$, with probability at least $1 - C_1\exp(-C_2\log p)$, we have
\bse
\|\wt\btheta - \btheta_0\|_1 \leq 16\lambda s_0/\kappa, \ \ \|\wt\btheta - \btheta_0\|_2   \leq \sqrt{32s_0}\lambda/\kappa ,  \  \ \text{and} \ \  \|(\X, \Z)(\btheta_0 - \wt\btheta)\|_2/\sqrt{n} \leq \lambda\sqrt{32s_0/\kappa},
\ese
where $\|\btheta_0\|_\infty \leq K_0$, $C^{\prime\prime}$ is a universal constant, $C_1$ and $C_2$ are
positive constants  depending on $K$, $K_{\epsilon}$, $K_U$, $K_0$,
$\kappa$ and $\sigma_U^2$ given in the proof,
$K_1 = 2K_U(K_0K+ K_{\epsilon})$, $K_2=4K(K+K_U)+2K_U^2+\sigma_U^2 $ and $K_3 = 4(K+K_U)^2 + 2\sigma_U^2$.
\end{Lemma}

The detailed proof is given in Appendix E.1 in supplementary materials. It is
based on the closeness condition for $\wh\bSig$ and $\wh\brho$, and
the restricted eigenvalue (RE) condition for matrix $\bSig$. Different
from deterministic design, Bernstein inequalities were used repeatedly
and we have shown that under the assumption that $s_0\sqrt{n^{-1}\log
  p}$ = o(1), the RE condition for sub-Gaussian matrix $\bSig$ holds
with probability at least $1-2p^{-\zeta}$ in Lemma F.2 in supplementary materials.

For the $l_{\infty}$ error bound, for simplicity,
slightly different notations are used here. Specifically,
we write $\btheta_0 = (\btheta_{0,S}\trans, \0\trans)\trans$, $(\X, \Z) = (\Q_S, \Q_{S^{C}})$,
and then partition the matrix $\bSig$ as
\bse
\bSig =
\begin{pmatrix}
&n^{-1}\Q_S\trans\Q_S &n^{-1}\Q_S\trans\Q_{S^C} \\
&n^{-1}\Q_{S^C}\trans\Q_S  &n^{-1}\Q_{S^C}\trans\Q_{S^C}
\end{pmatrix}
=
\begin{pmatrix}
\bSig_{S, S} &\bSig_{S, S^C} \\
\bSig_{S^C, S} &\bSig_{S^C, S^C}
\end{pmatrix}.
\ese
To clarify, the above partition is based on the true support of model
(\ref{eq:model}), that is, whether $\X$ is a part of $\Q_S$ depends on
the true value $\beta_0$.
Actually, when deriving the $l_\infty$ error bound for $\wt\btheta$,
whether $\beta_0$ equals $0$ would not affect the proof as well as the
theoretical result.
To derive the $l_{\infty}$ error bound for $\wt\btheta$, we need to further assume that
\be
\label{eq:condition_infty}
\lambda_{\min}\{E(\bSig_{S, S})\} = \kappa_S>0, \ \ \text{and} \ \ \|E(\bSig_{S^C, S})\{E(\bSig_{S, S})\}^{-1}\|_{\infty} \leq 1-\bg,
\ee
for some $\bg \in (0, 1]$. Let $\|E(\bSig_{S,S})^{-1}\|_{\infty} = \phi$ and $\|E(\bSig_{S,S})^{-1}\|_{\infty} = \Phi$.
The $l_{\infty}$ error bound result is stated as follows, which are similar to those given
in Theorem 2 in \cite{datta2017cocolasso} with minor
modifications.
The detailed proof is given in the Appendix E.2 in the supplementary materials.

\begin{Lemma}
\label{lemma:theta_infty}
Let $\lambda = C_{\lambda}s_0\sqrt{n^{-1}\log p}=o(1)$.
Under the assumptions given in (\ref{eq:condition_infty}) and $C_{\lambda} > 8K_4/(\gamma \sqrt{C^{\prime\prime}})$, where $K_4 = 2K^2K_0 + 2KK_{\epsilon}$
\begin{enumerate}
\item[(a)] With probability at least $1-p_1(\delta)$, there exists a unique solution $\wt\btheta$ minimizing
  $\btheta\trans\wt\bSig\btheta/2 - \wh\brho\trans\btheta +
  \lambda\|\btheta\|_1$ whose support is a subset of the true
  support.
\item[(b)]  With probability at least $1-p_2(\delta^{\prime})$, $\|\wt\btheta_S - \btheta_{0S}\|_{\infty}\leq C_{\infty}\lambda$, where $C_{\infty} = 8\phi$.
\end{enumerate}
Probabilities $p_1(\delta)$ and $p_2(\delta^{\prime})$ go to zero as $n$ goes to infinity and the detailed expressions are given in Appendix E.2 in supplementary materials.
\end{Lemma}

Note that Parts (a) and (b) of Lemma \ref{lemma:theta_infty} imply that under the given conditions, $\|\wt\btheta - \btheta_0\|_{\infty} = \|\wt\btheta_S - \btheta_{0S}\|_{\infty} \leq C_{\infty}\lambda$ with probability at least $1 - p_1(\delta) - p_2(\delta^{\prime})$.

\begin{Rem}
Note that we use $\wt\bSig$ in the loss function for CoCoLasso estimator to make the problem convex, but use $\wh\bSig$ in the loss function to construct decorrelated score function.
This discrepancy does not cause any problem when deriving
the theoretical properties of our corrected score test statistic and one-step estimator.
\end{Rem}

\begin{Rem}
For CoCoLasso estimator, the tuning parameter $\lambda$ has the order $s_0\sqrt{n^{-1}\log p}$. However, in \cite{loh2017support}, the tuning parameter has the order $\sqrt{n^{-1}\log p}$ under the assumption that $\|\btheta_0\|_1$ is bounded. In our proofs, we only assume that $\|\btheta_0\|_\infty$ is bounded. With the stronger assumption that $\|\btheta_0\|_1$ is bounded, the error bounds of CoCoLasso estimator would have the same order as those proposed in \cite{loh2017support} and \cite{belloni2017linear}.
\end{Rem}

\subsection{Algorithm}

Now we  summarize the proposed estimation procedure as the following algorithm.
\begin{enumerate}
\item

Calculate the initial CoCoLasso estimator $\wt\btheta = (\wt\beta,
\wt\bg\trans)\trans$.

\item

Estimate $\bomega$ by the Dantzig type estimator $\wh\bomega$,
\bse
\wh\bomega=\argmin \|\bomega\|_1 \ \ \text{s.t.}  \ \ \|\wh\bSig_{12}
- \bomega\trans\wh\bSig_{22}\|_{\infty}\leq \lambda^{\prime},
\ese
where $\lambda^{\prime} = O(\sqrt{\log p/n})$. For the detailed algorithm, see \cite{candes2007dantzig}.
Note that other penalized M-estimators can also be used to solve for $\wh \bomega$, for example, the Lasso.

\item

Calculate the estimated decorrelated score function
\bse
\wh S(\beta, \wt\bg) = S_{\beta}(\beta, \wt\bg) -
\wh\bomega\trans\S_{\bg}(\beta, \wt\bg),
\ese
and the test statistic
$\wh T_n = n^{1/2}\wh S(\beta^*, \wt\bg)(\wh\sigma^2_{\beta|\bg, H_0})^{-1/2}$,
where $\wh \sigma^2_{\beta|\bg, H_0}$ is given in (\ref{eq:sigmacond}).
Under the conditions given in Theorem \ref{th:asynormal1}, the test
statistic $\wh T_n$ is asymptotically standard normal.

\item Calculate the one-step estimator
\bse
\wh\beta &=& \wt\beta - \wh
  S(\wt\btheta)/\{\partial \wh S(\beta, \wt\bg)/\partial\beta\}|_{\beta
    = \wt\beta} \\
    &=& \wt\beta - \wh S(\wt\btheta)/(\wh\bSig_{11}- \wh\bomega\trans\wh\bSig_{21}).
\ese
Construct the $(1-\alpha)100\%$ confidence interval for $\beta_0$ as
 $[\wh\beta - z_{\alpha}\sqrt{\wh\sigma_\beta^2/n} ,  \wh\beta +
z_{\alpha}\sqrt{\wh\sigma_\beta^2/n}  ]$,
where $\Phi(z_{\alpha}) = 1-\alpha/2$, and $\wh\sigma_\beta^2$ is given in
Theorem \ref{th:asynormal2}.

\end{enumerate}

\section{Theory for Test and Confidence Intervals}
\label{sec:3}
%Similar to the strategy used in \cite{ning2017general}, 
We first establish
four technical lemmas \ref{lemma:consistency},
\ref{lemma:gradient_hessian}, \ref{lemma:smoothness} and \ref{lemma:clt} to ensure the asymptotic normality of the
corrected score test statistic $\wh
T_n$ and the one-step estimator $\wh\beta$.
Detailed descriptions of the four lemmas are given in Appendix A.

\subsection{Corrected Score Test}
\begin{Th}
\label{th:asynormal1}
Under conditions of  Lemmas \ref{lemma:consistency} - \ref{lemma:smoothness}
and under $H_0:\beta_0=\beta^*$, it follows that
\bse
n^{1/2}\wh S(\beta^*,\wt\bg) (\sigma^2_{\beta|\bg,0})^{-1/2} \to N(0,1)
\ese
in distribution.
\end{Th}
In Theorem \ref{th:asynormal1}, we state the asymptotic normality of the
decorrelated score test statistic by assuming its true variance is
known.
The detailed proof is given in Appendix
\ref{app:th:asynormal1}.
To show the asymptotic properties of the test statistic $\wh T_n$ with
estimated variance $\wh\sigma^2_{\beta\mid \bg, H_0}$, we need to
further study the difference between $\wh\sigma^2_{\beta\mid \bg,
  H_0}$ and $\sigma^2_{\beta|\bg,0}$, which is more complex than that
of linear models without measurement error.
We need to use $l_{\infty}$ error bound of $\wt\bg - \bg_0$ to facilitate the proof.
Under a stronger assumption that $s_0^3\sqrt{n^{-1}\log p} = o(1),$
we show that $\wh T_n$ is still asymptotically standard normal in the
following corollary and detailed proof can be found in Appendix
\ref{app:cor:scoretest}.

\begin{Cor}\label{cor:scoretest}
Suppose that $s_0^3\sqrt{n^{-1}\log p} = o(1)$. Under conditions of
Lemmas \ref{lemma:consistency} - \ref{lemma:smoothness} and under $H_0$, it follows that
\bse
n^{1/2}\wh S(\beta^*,\wt\bg) (\wh\sigma^2_{\beta|\bg,H_0})^{-1/2} \to N(0,1)
\ese
in distribution.
\end{Cor}

\begin{Rem}
Assume that $\log(p) = O(n^{a_1})$, $s_0 = O(n^{a_2})$ and
$s^{\prime} = O(n^{a_3})$. Then the conditions in Corollary
\ref{cor:scoretest} together with
$s_0(s_0\vee s^{\prime})n^{-1/2}\log p = o(1)$,
%and $s_0^3\sqrt{n^{-1}\log p} = o(1)$,
 imply that
\bse
a_2 + (a_2\vee a_3) + a_1 < 1/2\ \ \text{and} \ \
3a_2 + a_1/2 < 1/2.
\ese
The inference framework of \cite{ning2017general} requires   $(a_2 \vee a_3)
+ a_1 < 1/2$, while the consistency of CoCoLasso estimator of
 \cite{datta2017cocolasso} requires
$2a_2 + a_1/2 < 1/2$.
Our requirement on $(n, p, s_0, s^{\prime})$ here is stronger.
This is because the CoCoLasso estimator converges more slowly than
standard penalized M-estimators for high-dimensional linear models.
On the other hand, the inference framework based on decorrelation
operation needs stronger assumptions on dimensionality and sparsity
compared with pure estimation theory.
\end{Rem}

We further study the power of our test statistic $\wh T_n$ at local
alternatives in the following corollary, and its proof is given in
Appendix \ref{app:altpower}.

\begin{Cor}\label{cor:scoretest_alt}
Consider the local alternative
$\beta_n = \beta^* + h/\sqrt{n} $,
where $h$ is a constant.
Under the assumptions given in Corollary \ref{cor:scoretest},
our score test statistic $\wh T_n = n^{1/2}\wh S(\beta^*,
\wt\bg)(\wh\sigma^2_{\beta|\bg, H_0})^{-1/2}$ converges to
$N\{-h(\sigma_{\beta_n}^2)^{-1/2}, 1\}$ in distribution under the
local alternatives, where $\sigma_{\beta_n}^2= [E\{\partial S(\beta,
\bg_0)/\partial\beta \mid_{\beta =
  \beta_n}\}]^{-2}\sigma^2_{\beta_n|\bg, 0}$, and
$\sigma^2_{\beta_n|\bg, 0}$ is $\sigma^2_{\beta|\bg, 0}$ with
$\beta_0$ replaced by $\beta_n$.
\end{Cor}

\subsection{Confidence Interval}
In addition to hypothesis testing, we also  construct asymptotic
confidence intervals for the parameter of interest $\beta$ based on
the one-step estimator $\wh\beta$. Its asymptotic normality is given
in the following theorem and the detailed proof is given in Appendix
\ref{app:th:asynormal2}.

\begin{Th}
\label{th:asynormal2}
Suppose conditions of
Lemmas \ref{lemma:consistency} - \ref{lemma:clt} are valid,
if $E[\{\partial S(\beta, \bg_0)/\partial \beta\}|_{\beta = \beta_0}]
\geq C$ for some positive constant $C$,
then
\bse
n^{1/2}(\wh\beta - \beta_0) =-\left[E\left\{
    \frac{\partial S(\beta,\bg_0)}{\partial
      \beta}\bigg\arrowvert_{\beta =
      \beta_0}\right\}\right]^{-1}n^{1/2}S(\beta_0, \bg_0)+o_P(1) \to
N(0, \sigma_{\beta}^2)
\ese
in distribution,
where the asymptotic variance
$
\sigma_\beta^2 =\{E(X_i^2)-\bomega\trans E(X_i\Z_i)\}^{-2}\sigma_{\beta\mid\bg,0}^2.
$
The variance $\sigma_\beta^2$ can be estimated as
\be
\label{eq:asyvariance}
\wh\sigma_\beta^2 =
\left(1-\wh\bomega\trans\wh\bSig_{21}\right)^{-2}
\left\{
(\wh\sigma_\epsilon^2+\wh\beta^2\sigma_U^2)(1-\wh\bomega\trans\wh\bSig_{21})+\wh\beta^2E(U_i^4)+\wh\sigma_\epsilon^2\sigma_U^2-\wh\beta^2\sigma_U^4
\right\},
\ee
where $\wh\sigma_\epsilon^2 =  n^{-1}\sum_{i=1}^n(Y_i - \wh\beta
W_i - \wt\bg\trans\Z_i)^2  -
\wh\beta^2\sigma_U^2$.
\end{Th}

\begin{Rem}
Lemma \ref{lemma:theta_infty} shows that the
sign consistency property of the CoCoLasso estimator is ensured by the
minimal signal condition $\min_{j\in S}
|\btheta_j|>C_{\infty}\lambda$.
That is, when $|\beta_0| <C_{\infty}\lambda$, then the CoCoLasso
estimate $\wt\beta$ will be set to $0$ with high probability.
With the decorrelation operation,
the convergence performance of our one-step estimator $\wh\beta$ is
improved significantly.
Meanwhile, our test statistic $\wh T_n$ retains power under the local
alternatives around $0$.
\end{Rem}

\begin{Rem}
In low dimensional case, 	
\cite{nakamura1990corrected} provided inference results of generalized
linear models with measurement error using corrected score functions.
We have established inference results in high-dimensional settings.
Since  $\sigma_{\beta\mid \bg}^2$ is the variance of the corrected
decorrelated score $S(\beta, \bg)$,  
the form of our asymptotic variance $\sigma_\bb^2$ is similar to theirs.
Further, we show that our one-step estimator is semiparametrically efficient. 
The extension to generalized linear models is important but beyond the scope of this paper.
\end{Rem}

\section{Empirical Studies}
\label{sec:4}
\subsection{Simulation Studies}

We conducted simulation studies under
different settings to investigate 
the performance of our proposed corrected decorrelated score test  and
the one-step estimator.  The code is available for public use.
To generate the data matrix $(\X, \Z)$, we simulated $n=100$ and $n=200$ independent
and identically distributed samples from a multivariate Gaussian
distribution $N_p(\0,\bSig)$, where $p = 250$ and $\bSig$ is the
autoregressive matrix with its entry $\bSig_{jk} = \rho^{|j-k|}$.
We considered two cases, where $\rho=0.25$ and $\rho=0.5$. To generate
the responses $\Y$, we added the
 regression error $\beps$ following the normal distribution $N(\0,
\sigma_{\epsilon}^2\I_n)$, where $\sigma_{\epsilon} = 0.2$.
The measurement error $\U$ was generated from $N(\0, \sigma_U^2 \I_n)$.
Three different values of $\sigma_U$ are considered, where
$\sigma_U=0.1, 0.15$ and 0.2 respectively.
Both estimation and inference  become progressively
more difficult with larger measurement error variance.
We considered two scenarios
for the true parameter $\btheta_0 = (\beta_0, \bg_0\trans)\trans$.
In the first scenario, $\btheta_0  = (1, 1, 0,\dots 0)\trans$. In the
second scenario, we set $\btheta_0  = (1, 0.8, 1.5, 0,\dots 0)\trans$.
Our goal is to test $H_0: \beta_0 = 1$ versus $H_1: \beta_0 \neq 1$.

For the initial CoCoLasso estimator $\wt\btheta$, we first perform variable
selection using (\ref{eqn:CCL}).
%with $\lambda = 2\sqrt{\log p/n}$.
Then refit the model using the selected covariates and set the
coefficients of the rest of the covariates to zero. During the procedure,
the tuning parameter $\lambda^{\prime}$ in (\ref{eq:dantzig}) is
selected by a K-fold cross-validation, where $K=4$.
Specifically, the optimal $\lambda^{\prime}$ is chosen in the sense of
$l_2$ prediction for the test sample, see \cite{bickel2007}.

In each setting, 1000 simulations are conducted.
The averaged type I error rates at significance levels $\alpha = 1\%,
5\%$ and $10\%$ of our test are summarized in Table \ref{t1}.
We can see that the type I error rates are very close to the nominal
significance levels in all the simulation settings.
To examine the power of our test, we regenerated data with $\beta_0
= 1.05, 1.10, 1.15$ and report the rejection rate at different
significance levels ranging from $1\%$ to $10\%$.
The results, together with the rejection rates under
$H_0$ when $\beta_0 = \beta^*
=1$, are shown in Figure \ref{fig:1}, as well as Figures S1, S2 and S3  in supplementary materials.
Overall, the test
has very good performance in terms of level under $H_0$, reflected in
the close approximation of the observed rejection rates and the
nominal levels. The power performance is also satisfactory in general,
where the curves representing the rejection rates
under all three alternatives are well separated from the null
rejection curve, and the power increases when sample size increases,
 the correlation $\rho$ decreases, the nonzero covariates number is smaller, or the
 measurement error variance decreases.

We also provide the performance of our one-step estimator $\wh\beta$
in Table 2, where we report
the mean and standard deviation of 1000 estimates of $\wh\beta$, as
well as the average of
the estimated asymptotic standard deviation calculated based on
(\ref{eq:asyvariance}).
In addition, we constructed the $95\%$ confidence intervals in each
simulation using the asymptotic normality of $\wh\beta$, and computed the
empirical coverage of the true value $\beta_0$.
We find that the one-step estimator performs well in different simulations settings.
In each setting,
the difference between the mean of the estimates and the true value
is very small, the mean of estimated standard deviations closely
approximates the empirical value, and the empirical coverage of the
estimated $95\%$ confidence intervals is reasonably close to the nominal level.

We have assumed $\sigma_U^2$ and $E(U_i^4)$ to be known. In this
section, we further conducted simulation studies to examine the impact
of $\wh \sigma_U^2$ and $\wh E(U_i^4)$. The simulation results are in the
supplementary materials H.2.

\begin{table}[t!]
% TABLE 1
\centering
\caption{Type I error of the corrected decorrelated score test at different significance levels}
\label{t1}
\begin{tabular}{cccccccccc}
\hline

 & &\multicolumn{4}{c}{ Scenario 1 } &\multicolumn{4}{c}{Scenario 2}    \\
          \cmidrule(lr){3-6}     \cmidrule(lr){7-10}
 & &\multicolumn{2}{c}{$\rho=0.25$}  &\multicolumn{2}{c}{$\rho = 0.5$} &\multicolumn{2}{c}{$\rho = 0.25$}  &\multicolumn{2}{c}{$\rho = 0.5$}  \\
          \cmidrule(lr){3-4}     \cmidrule(lr){5-6}  \cmidrule(lr){7-8}     \cmidrule(lr){9-10}
 $\sigma_U$ &$\alpha$  & \multicolumn{1}{c}{$n=100$}  &\multicolumn{1}{c}{$n=200$}   &\multicolumn{1}{c}{$n=100$}  &\multicolumn{1}{c}{$n=200$} &\multicolumn{1}{c}{$n=100$}  &\multicolumn{1}{c}{$n=200$} &\multicolumn{1}{c}{$n=100$}  &\multicolumn{1}{c}{$n=200$} \\
\hline
0.1&$1\%$     &$1.1\%$ &$0.9\%$  &$1.6\%$ &$1.1\%$  &$1.0\%$   &$0.8\%$   &$1.4\%$ &$1.1\%$\\
     &$5\%$     &$5.6\%$ &$4.4\%$  &$5.2\%$ &$5.4\%$  &$5.6\%$   &$4.4\%$   &$5.5\%$ &$5.5\%$\\
     &$10\%$   &$9.8\%$ &$10.6\%$&$9.4\%$ &$12.0\%$&$10.2\%$ &$10.7\%$ &$9.3\%$ &$12.0\%$ \\

0.15&$1\%$    &$1.4\%$   &$1.0\%$   &$1.3\%$ &$1.1\%$   &$1.1\%$ &$1.1\%$   &$1.4\%$   &$0.9\%$\\
       &$5\%$    &$5.4\%$   &$4.9\%$   &$5.4\%$ &$5.9\%$   &$4.6\%$ &$5.4\%$   &$5.9\%$   &$5.9\%$ \\
       &$10\%$  &$11.3\%$ &$10.8\%$ &$9.9\%$ &$11.4\%$ &$9.2\%$ &$10.9\%$ &$10.5\%$ &$12.0\%$\\

0.2&$1\%$     &$1.5\%$   &$1.0\%$   &$1.4\%$   &$0.8\%$   &$2.1\%$  &$1.2\%$   &$1.5\%$   &$0.7\%$\\
     &$5\%$     &$6.2\%$   &$5.9\%$   &$5.9\%$   &$5.7\%$   &$6.0\%$  &$6.3\%$   &$6.4\%$   &$5.7\%$ \\
     &$10\%$   &$10.9\%$ &$10.9\%$ &$10.7\%$ &$11.7\%$ &$11.1\%$ &$10.5\%$ &$11.3\%$ &$11.3\%$\\
\hline
\end{tabular}
\end{table}

\begin{figure}[t!]
    \centering
    \includegraphics[width=1\textwidth]{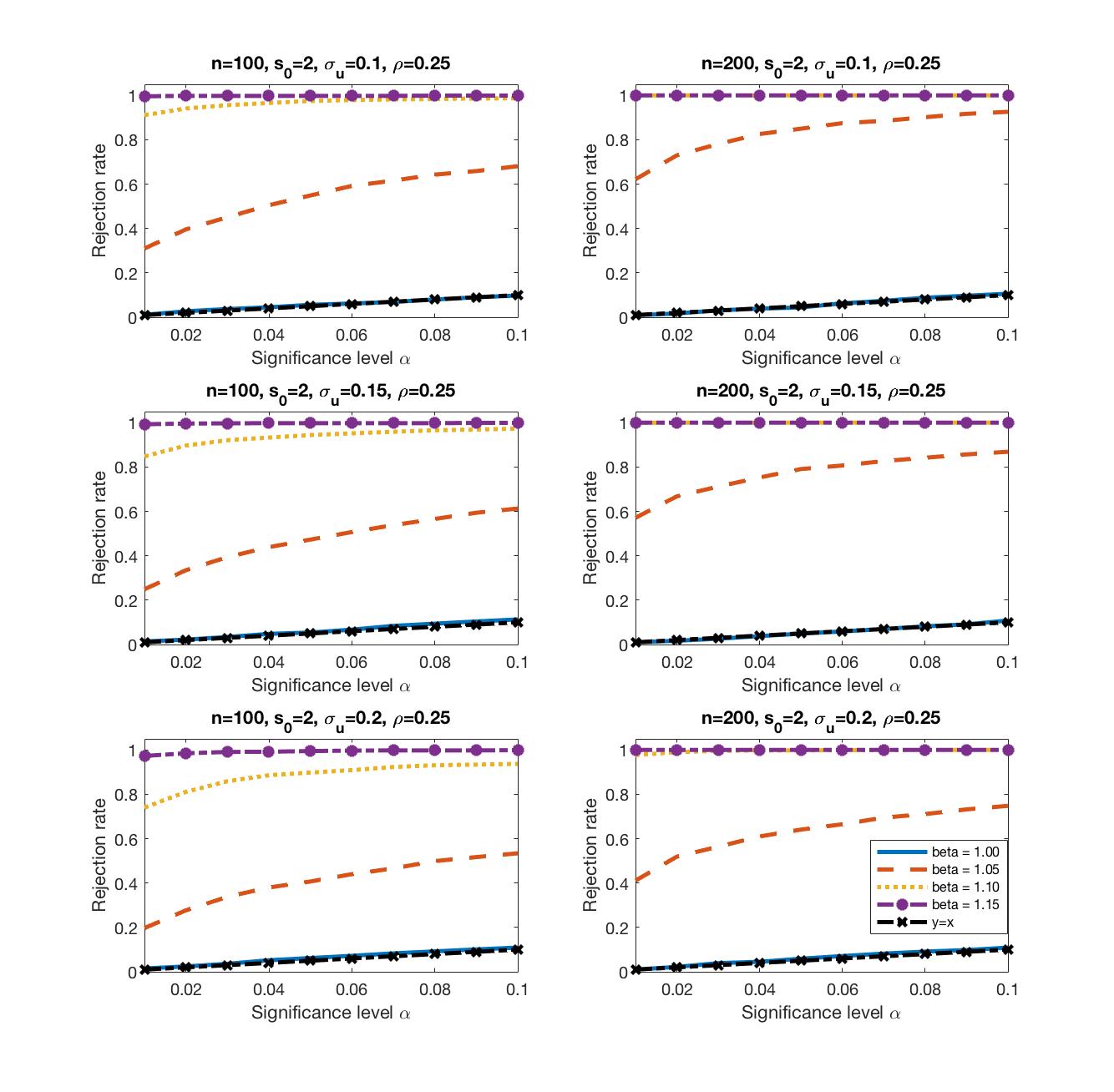}
    \caption{Power of the proposed corrected decorrelated score test at different significance levels in scenario 1 with $\rho = 0.25$}
\label{fig:1}
\end{figure}

\begin{table}[t!]
% TABLE 2
\centering
\begin{threeparttable}
\caption{Performance of the one-step estimator $\wh\beta$}
\label{t2}
\begin{tabular}{clcccccccc}
\hline

& &\multicolumn{4}{c}{ Scenario 1 } &\multicolumn{4}{c}{Scenario 2}    \\
\cmidrule(lr){3-6}     \cmidrule(lr){7-10}
& &\multicolumn{2}{c}{$\rho=0.25$}  &\multicolumn{2}{c}{$\rho = 0.5$} &\multicolumn{2}{c}{$\rho = 0.25$}  &\multicolumn{2}{c}{$\rho = 0.5$}  \\
\cmidrule(lr){3-4}     \cmidrule(lr){5-6}  \cmidrule(lr){7-8}     \cmidrule(lr){9-10}
$\sigma_U$&  & \multicolumn{1}{c}{$n=100$}  &\multicolumn{1}{c}{$n=200$}   &\multicolumn{1}{c}{$n=100$}  &\multicolumn{1}{c}{$n=200$} &\multicolumn{1}{c}{$n=100$}  &\multicolumn{1}{c}{$n=200$} &\multicolumn{1}{c}{$n=100$}  &\multicolumn{1}{c}{$n=200$} \\
\hline
0.1& Mean &1.0008 &1.0000 &1.0010 &1.0001 &1.0008 &1.0000 &1.0003 &0.9996 \\
&Est sd       &0.0237 &0.0163 &0.0259 &0.0177 &0.0235 &0.0162 &0.0257 &0.0177 \\
&Emp sd     &0.0246 &0.0167 &0.0259 &0.0185 &0.0246 &0.0167 &0.0260 &0.0185 \\
&Emp cvg    &$94.1\%$ &$94.6\%$ &$93.6\%$ &$93.8\%$ &$93.9\%$ &$94.7\%$ &$93.7\%$ &$93.6\%$ \\

0.15& Mean &1.0024 &1.0011 &1.0031 &1.0012 &1.0024 &1.0011  &1.0028 &1.0009\\
&Est sd      &0.0268 &0.0183 &0.0294 &0.0199 &0.0267 &0.0183  &0.0292 &0.0199 \\
&Emp sd    &0.0279 &0.0183 & 0.0299 &0.0204 &0.0270 &0.0184 &0.0302 & 0.0204\\
&Emp cvg  &$94.2\%$ &$94.9\%$  &$93.3\%$ &$94.0\%$ &$94.6\%$ &$94.7\%$ &$93.4\%$ &$93.7\%$ \\

0.2& Mean  &1.0049 &1.0015 &1.0062 &1.0024 &1.0085 &1.0016 &1.0061 &1.0021 \\
&Est sd       &0.0309 &0.0211 &0.0341 &0.0230 &0.0313 &0.0211 &0.0339 &0.0229 \\
&Emp var   &0.0324 &0.0217 &0.0355 &0.0234 &0.0333 &0.0218 &0.0359 &0.0234 \\
&Emp cvg  &$94.1\%$ &$94.1\%$ &$93.0\%$ &$93.5\%$ &$92.9\%$ &$93.9\%$ &$92.2\%$ &$93.2\%$  \\
\hline
\end{tabular}
\begin{tablenotes}
\item In Table \ref{t2},  ``Est sd'' denotes the mean of 1000 estimated asymptotic
standard deviations; ``Emp sd'' denotes the empirical
standard deviation of 1000 estimates; 
``Emp cvg" denotes the empirical coverage of the estimated $95\%$ CI for $\bb_0$.
\end{tablenotes}
\end{threeparttable}
\end{table}

\subsection{Real Data Analysis}
\label{subsec:realdata}

We illustrate the proposed procedure via an empirical analysis of
a data set analyzed in \cite{chu2016feature}. The data set was collected in
%The Childhood Asthma Management Program
%is
a clinical trial designed to determine the long-term effects of
different inhaled treatments for mild to moderate childhood asthma,
where phenotypic information and genome-wide SNP data are accessible.
%from dbGap study accession phs000166.v2.p1.
The FEV1/FVC ratio is an important index used in diagnosis of
obstructive and restrictive lung disease, which represents the
proportion of a person's vital capability to expire in the first
second of forced expiration to the full vital capacity.
We are interested in understanding how this ratio, often measured with
errors, together with
basic demographic variables and SNPs would affect the severity of
asthma symptoms in children.

Here we focus on $n=199$ subjects in the
nedocromil treatment group, each had four clinical visits over 8 months.
Exploratory data analysis was conducted on the four measurements of
FEV1/FVC ratio, and no visible time trend was detected.
The response variable $Y_i$ is the average asthma symptoms (amsys).
We let $X_i$ be the unobserved FEV1/FVC ratio and $W_i$ be the average
of four measurements with homoscedastic measurement errors.
Standard deviation and the fourth moment of measurement error $U_i$
are estimated using the four measurements for each subject based on
the fact that
$W_{ik} - W_{ij} = U_{ik} - U_{ij}$, $\var(U_{ik} - U_{ij}) =
2\sigma_U^2$, and $E(U_i^4) = [E\{( U_{ik} - U_{ij})^4\} -
6\sigma_U^4]/2$ for $i=1,\dots n$ and $j,k = 1,\dots, 4$.
Note that we do not need to assume the normality of measurement errors here.
The estimated values are $\wh\sigma_U = 0.4625$ and $\wh E(U_i^4) = 0.1719$.
The error-free variables $\Z_i$ are gender, age at baseline and 676
SNPs screened based on minor allele frequency (MAF).
Here we treat SNPs as continuous variables by assuming that having two
of the minor alleles has twice the effect on the phenotype as having
one of the minor alleles, and zero means no effect.

Our goal is to first select significant variables among $p = 679$
variables in model (\ref{eq:model}), estimate the corresponding
coefficients and then make inference for the error-prone variable
FEV1/FVC ratio based on the proposed corrected decorrelated score
test and the asymptotic properties of the one-step estimator.
For the initial CoCoLasso estimator $\wt\btheta$,  the tuning
parameter $\lambda$ is selected by cross validation with the criterion
proposed in \cite{datta2017cocolasso}.
We find that
besides FEV1/FVC ratio which is of interest, seven SNPs are
selected. Detailed information about the selected SNPs is given in
Table \ref{t3}.

\begin{table}[t!]
    \centering
    \caption{Information about the seven SNPs selected by CoCoLasso method}
    \label{t3}
    \begin{tabular}{lccccccc}
    \hline
         &${\rm SNP}_1$ &${\rm SNP}_2$ &${\rm SNP}_3$ &${\rm SNP}_4$ &${ \rm SNP}_5$ &${\rm SNP}_6$ &${\rm SNP}_7$  \\
         \hline
         SNP name &rs2830066 &rs11798747 &rs6961655 &rs4432291 &rs6860832 &rs699770 &rs4520841\\
         Chromosome &21 &X &7 &17 &5 &1 &16 \\
         Chr.position &26121885 &18889776 &136422490 &72610903 &8451644 &119318352 &26088794\\
         Coefficient &0.0143 &-0.0125 &-0.0118 &-0.0092 &-0.0056 &-0.0046 &-0.0025\\
    \hline
    \end{tabular}
\end{table}

Under the null hypothesis $H_0: \beta_0 = 0$, the corrected
decorrelated score test statistic $\wh T_n = 4.9806$. Hence, we
reject the null hypothesis.
The CoCoLasso estimate for $\beta$ is $-0.0654$, while the one-step
estimate is $-0.1101$ with confidence interval $(-0.1508, -0.0693)$.
The negativeness of $\wh\beta$ verifies the fact that the lower the
FEV1/FVC ratio, the severer the obstruction of air escaping from the
lungs.

Throughout the data analysis,
we estimated the second and fourth moments of the measurement
error using the four measurements of each subject.  
Because of the independent error assumption, 
$U_{ik} + U_{ij}$ is  uncorrelated to  $U_{ij} - U_{ik}$. 
Recall that the $W_i$ relies on $U_{ik} + U_{ij}$, while the error moment
estimates are  based on $U_{ij} - U_{ik}$. Under normality assumption,
the standard errors of the two moment estimates do not affect the
performance of our proposed inference procedure.

\section{Discussion}
In this paper, we have proposed an inference procedure for high-dimensional linear measurement error models based on corrected decorrelated score functions.
With the decorrelation operation,
our corrected score test statistic $\wh T_n$
is asymptotically normal and retains power under the local alternatives around 0. 
Further, the convergence rate of the one-step estimator $\wh \bb$ has
significantly improved compared to that of the initial
  estimator and achieves the semiparametric efficiency.
Here we have assumed that the variance and the fourth moments of the
measurement error are known.  
The framework in this paper still works if we  treat $\sigma_U^2$ and
$E(U_i^4)$ as nuisance parameters and then conduct decorrelation.  
Specifically, the new nuisance parameters are $(\bg\trans, \sigma_U^2,
E(U_i^4))$. Note that we do not impose any penalty on $\sigma_U^2$ and
$E(U_i^4)$.

One further research direction is to develop inference procedures when the number of covariates with measurement errors diverges with sample size $n$.
Another possible consideration is to relax the sparsity assumption on $\bomega$. That is, extend the theory to cases where the ordered entries of $\bomega$ decay at a certain rate.

\section*{Appendix A: Four technical Conditions}

%\renewcommand{\thesection}{A.\arabic{section}}

%\setcounter{equation}{0}
%\renewcommand{\theequation}{A.\arabic{equation}}

%\setcounter{lemma}{0}
%\renewcommand{\thelemma}{A\arabic{lemma}}

%\setcounter{Cor}{0}
%\renewcommand{\theCor}{D.\arabic{Cor}}
%\setcounter{subsection}{0}
%\renewcommand{\thesubsection}{A.\arabic{subsection}}
%Following \cite{ning2017general}, we establish four technical lemmas \ref{lemma:consistency}, \ref{lemma:gradient_hessian}, \ref{lemma:smoothness} and \ref{lemma:clt} to ensure the asymptotic normality of the corrected decorrelated score test statistic $\wh T_n$ and the one-step estimator $\wh\beta$.

    \begin{Lemma}
        \label{lemma:consistency}
        Recall that $S^{\prime}=\rm{supp}(\bomega)$ and  $|S^{\prime}| = s^{\prime}$. Let $\lambda^{\prime}= C_{\lambda^{\prime}} \sqrt{n^{-1}\log p}$.
        The Dantzig type estimator $\wh\bomega$ satisfies
        $\|\wh\bomega - \bomega\|_1 = O_P(s^{\prime}\sqrt{n^{-1}\log{p}})$,
        when $C_{\lambda^{\prime}}>\sqrt{2K_5/C^{\prime\prime}}$.
        Here $C^{\prime\prime}$ is a universal constant
        and $K_5 = 2K(K+K_U+KK_{\omega})$.
    \end{Lemma}

    \begin{Lemma}
        \label{lemma:gradient_hessian}
        Let $\bnu = (1, -\bomega\trans)\trans$.
        The gradient and Hessian of the corrected loss function (\ref{eq:correctedloss}) satisfy
        $\|\S_{\btheta}(\btheta_0)\|_{\infty} = O_p(\sqrt{n^{-1}\log{p}})$ and
        $\|\bnu\trans\nabla \S_{\btheta}(\btheta_0) -
        E\{\bnu\trans\nabla \S_{\btheta} (\btheta_0)\}
        \|_{\infty} = O_P(\sqrt{n^{-1}\log{p}})$.
    \end{Lemma}

    \begin{Lemma}
        \label{lemma:smoothness}
        Let $\wt\btheta_{H_0} = (\beta^*, \wt\bg\trans)\trans$, $\wh\bnu = (1, -\wh\bomega\trans)\trans$.
        Assume that
        \bse
        \frac{s_0(s^{\prime} \vee s_0) \log p} {\sqrt{n}} = o(1).
        \ese
        Then
        $\bnu\trans\{\S_{\btheta}(\widecheck{\btheta}) -
        \S_{\btheta}(\btheta_0) - \nabla
        \S_{\btheta}(\btheta_0)(\widecheck{\btheta} - \btheta_0)\} =0$,
        and $(\wh\bnu-\bnu)\trans\{\S_{\btheta}(\widecheck{\btheta}) - \S_{\btheta}(\btheta_0)\}=o_P(n^{-1/2})$,
        for both $\widecheck{\btheta} =
        \wt\btheta_{H_0}$ and $\widecheck{\btheta} = \wt\btheta$.
    \end{Lemma}

    \begin{Lemma}
        \label{lemma:clt}
        When  (\ref{eq:model}) does not degenerate, i.e., the corrected
        decorrelated score function $S(\btheta)\not\equiv 0$ a.s., then
        \bse
        \sqrt{n}\bnu\trans \S_{\btheta}(\btheta_0)(\sigma^2_{\beta|\bg, 0})^{-1/2} \to N(0,1)
        \ese
        in distribution.
        Here $\sigma^2_{\beta|\bg, 0} = (\sigma_{\epsilon}^2 +
        \beta_0^2\sigma_U^2)\{1 - \bomega\trans E(X_i\Z_i)\} + \beta_0^2
        E(U_i^4) +\sigma_{\epsilon}^2\sigma_U^2 - \beta_0^2\sigma_U^4$
        by (\ref{eq:matrixI}) and (\ref{eq:sigmaS}), and $\sigma^2_{\beta|\bg, 0}  \geq C$ for some positive constant $C$.
    \end{Lemma}

Lemma \ref{lemma:consistency}, together with Lemma \ref{lemma:theta},
states the consistency properties for initial estimators $\wt\btheta$
and $\wh\bomega$, which are crucial to the asymptotic performance of
our corrected test statistic and one-step estimator.
Lemma \ref{lemma:gradient_hessian} and Lemma \ref{lemma:smoothness}
describe the concentration properties of the gradient and Hessian of the
corrected loss function (\ref{eq:correctedloss}), and its local
smoothness properties, respectively.
For high-dimensional random designs, it is important to quantify the
distance between sample level statistic and its corresponding
population level value, especially for critical statistics like the score
function and the Hessian matrix.
For local smoothness, \cite{ning2017general} require $(s_0
\vee s^{\prime})n^{-1/2}\log p = o(1)$.
However, using CoCoLasso estimator as the initial estimator,
we need a stronger condition on dimensionality and sparsity to
guarantee the $n^{-1/2}$ rate local smoothness of the corrected loss
function.
Lemma \ref{lemma:clt} is the central limit theorem for corrected
decorrelated sore function $S(\btheta_0)$, which is a linear
combination of $S_{\btheta}(\btheta_0)$.
Because we define the score function as the gradient of the corrected
loss function, which is different from negative log-likelihood,
the variance $\sigma^2_{\beta\mid\bg,0}$ of $S(\btheta_0)$ has relatively complex form.
Detailed proofs of the four lemmas are given in Appendices F.1,
F.2, F.3 and F.4, respectively, in supplementary materials.

\section*{Appendix B: Proofs Regarding  Score Test Statistic}

\setcounter{subsection}{0}\renewcommand{\thesubsection}{B.\arabic{subsection}}

\subsection{Proof of Theorem \ref{th:asynormal1}}
\label{app:th:asynormal1}
\begin{proof}
Recall that $\wt\btheta_{H_0} = (\beta^*, \wt\bg\trans)\trans$, $\wh\bnu =
(1, -\wh\bomega\trans)\trans$ and $\bnu = (1, -\bomega\trans)\trans$.
We have
\bse
&&\sqrt{n} |\wh S(\wt\btheta_{H_0}) - S(\btheta_0)| \\
&=&\sqrt{n}|\wh \bnu\trans \S_{\btheta}(\wt\btheta_{H_0}) - \bnu\trans \S_{\btheta}(\btheta_0)| \\
&\leq& \sqrt{n} |\bnu\trans\{\S_{\btheta}(\wt\btheta_{H_0}) -
S_{\btheta}(\btheta_0)\}| + \sqrt{n} |(\wh\bnu -
\bnu)\trans\S_{\btheta}(\wt\btheta_{H_0})| \\
&=&D_1 + D_2,
\ese
where $D_1\equiv\sqrt{n} |\bnu\trans\{\S_{\btheta}(\wt\btheta_{H_0}) -
S_{\btheta}(\btheta_0)\}|$ and $D_2\equiv \sqrt{n} |(\wh\bnu -
\bnu)\trans\S_{\btheta}(\wt\btheta_{H_0})|$.
Since the corrected loss function (\ref{eq:correctedloss}) is a quadratic function of $\btheta$,
by Lemma \ref{lemma:smoothness}, we have
\be
\label{eq:th1}
|D_1|
&=&\sqrt{n} |\bnu\trans \nabla \S_{\btheta}(\btheta_0)
(\wt\btheta_{H_0} - \btheta_0)|\n \\
&=& \sqrt{n} |\bnu\trans \nabla \S_{\btheta}(\btheta_0) (0, \wt\bg\trans - \bg_0\trans)\trans| \n \\
&\leq& \sqrt{n}\|\wt\bg - \bg_0\|_1\|\wh\bSig_{12} - \bomega\trans\wh\bSig_{22} \|_{\infty}\n \\
& = & \sqrt{n} \|\wt\btheta_{H_0} -
\btheta_0\|_1\|\wh\bSig_{12} - \bomega\trans\wh\bSig_{22} \|_{\infty} \n \\
& = & \sqrt{n} \|\wt\btheta_{H_0} -
\btheta_0\|_1\| \wh\bSig_{12} - \bomega\trans\wh\bSig_{22} -
E\{\wh\bSig_{12} - \bomega\trans\wh\bSig_{22}\}
\|_{\infty} \n \\
&&+\sqrt{n}
\|\wt\btheta_{H_0} -
\btheta_0\|_1\|E\{ \wh\bSig_{12} - \bomega\trans\wh\bSig_{22}\}\|_{\infty} \n \\
& = & \sqrt{n} \|\wt\btheta_{H_0} -
\btheta_0\|_1\| \wh\bSig_{12} - \bomega\trans\wh\bSig_{22} -
E\{\wh\bSig_{12} - \bomega\trans\wh\bSig_{22}\}
\|_{\infty} \n  \\
& \le & \sqrt{n} \|\wt\btheta_{H_0} -
\btheta_0\|_1
\|\bnu\trans \nabla \S_{\btheta}(\btheta_0)
-E\{\bnu\trans \nabla \S_{\btheta}(\btheta_0)\}
 \|_{\infty}.
\ee
In the above derivation, we used the fact that
$\nabla \S_{\btheta}(\theta_0) =
\wh\bSig$, and
under $H_0$ the first element of
$\wt\btheta_{H_0} - \btheta_0$ is $0$. In addition,
$\bomega\trans =
E(W_i\Z_i\trans)E(\Z_i\Z_i\trans)^{-1}$, and hence
\bse
E(\wh\bSig_{12} - \bomega\trans\wh\bSig_{22})
=E(W_i\Z_i\trans) - \bomega\trans E(\Z_i\Z_i\trans)
=\0.
\ese
By Lemmas \ref{lemma:theta} and \ref{lemma:gradient_hessian}, we
have $D_1 \leq O_P\left\{ s_0^2\sqrt{n^{-1}\log p}\cdot\sqrt{\log p}\right\} = o_P(1)$.

For $D_2$, Lemma \ref{lemma:smoothness} yields
\bse
|D_2|
&\le&\sqrt{n} |(\wh\bnu -
\bnu)\trans\S_{\btheta}(\btheta_0)|+
\sqrt{n} |(\wh\bnu -
\bnu)\trans\{\S_{\btheta}(\wt\btheta_{H_0})-\S_{\btheta}(\btheta_0)\}|\\
&\leq& \sqrt{n} |(\wh\bnu - \bnu)\trans \S_{\btheta}(\btheta_0)|  + o_P(1) \\
&\leq& \sqrt{n} \|\wh\bnu -\bnu\|_1 \|\S_{\btheta}(\btheta_0)\|_{\infty} + o_P(1).
\ese
By Lemmas \ref{lemma:consistency} and \ref{lemma:gradient_hessian}, we
have $|D_2|\leq O_P\left\{s^{\prime}\sqrt{n^{-1}\log p}\cdot\sqrt{\log p}\right\}+o_P(1) = o_P(1)$.
Hence, we have $ \sqrt{n}|\wh S(\wt\btheta_{H_0}) - S(\btheta_0)|=o_P(1)$.
Since $\sigma^2_{\beta|\bg,0} >0$, we obtain that
\bse
\sqrt{n}\mid\wh S(\wt\btheta_{H_0}) (\sigma^2_{\beta|\bg,0})^{-1/2} - S(\btheta_0)
  (\sigma^2_{\beta|\bg,0})^{-1/2}\mid=o_P(1).
\ese
By Lemma \ref{lemma:clt},
$\sqrt{n}S(\btheta_0)(\sigma^2_{\beta|\bg,0})^{-1/2}
=\sqrt{n}\bnu\trans \S_\btheta(\btheta_0)(\sigma^2_{\beta|\bg,0})^{-1/2}
\to N(0,1)$ in distribution.
Applying the Slutsky's theorem, we hence get
$\sqrt{n}\wh S(\beta^*, \wt\bg)(\sigma^2_{\beta|\bg,0})^{-1/2} \to
N(0,1)$ in distribution under null hypothesis. This
completes the proof.
\end{proof}

\subsection{Proof of Corollary \ref{cor:scoretest}}
\label{app:cor:scoretest}
\begin{proof}
Recall that $\wh T_n = n^{1/2}\wh S(\beta^*, \wt\bg)(\wh\sigma_{\beta\mid\bg, H_0}^2)^{-1/2}$.
Let $T_n = n^{1/2}\wh S(\beta^*, \wt\bg)(\sigma_{\beta\mid\bg, 0}^2)^{-1/2}$.
Then
\bse
\wh T_n - T_n  = T_n \left( \frac{\sigma_{\beta\mid\bg, 0}}{\wh\sigma_{\beta\mid\bg, H_0}}-1 \right).
\ese
In Theorem \ref{th:asynormal1}, we have proved that $T_n \to N(0,1)$ in distribution, as $n\to\infty$.
It remains to show that
$\sigma_{\beta\mid\bg,0}/\wh\sigma_{\beta\mid\bg, H_0}-1 =
o_p(1)$.
We start from deriving the bound of $|\wh\sigma_{\beta\mid\bg, H_0}^2 - \sigma_{\beta\mid\bg,0}^2 |$.
Recall that
\bse
\wh\sigma_{\beta\mid\bg, H_0}^2 &=& (\wt\sigma_{\epsilon, H_0}^2 + \beta^{*2}\sigma_U^2)(1 -
\wh\bomega\trans\wh\bSig_{21}) + \beta^{*2}E(U_i^4) +
\wt\sigma_{\epsilon, H_0}^2\sigma_U^2 - \beta^{*2} \sigma_U^4, \\
\sigma_{\beta\mid\bg,0}^2 &=& (\sigma_{\epsilon}^2 + \beta_0^2\sigma_U^2)\{1 -
\bomega\trans E(X_i\Z_i)\} + \beta_0^2E(U_i^4) +
\sigma_{\epsilon}^2\sigma_U^2 - \beta_0^2 \sigma_U^4.
\ese
Since $\beta_0 = \beta^*$ under null hypothesis,
then we have
\bse
\wh\sigma_{\beta\mid\bg, H_0}^2  - \sigma_{\beta\mid\bg,0}^2
&=&  (\wt\sigma_{\epsilon, H_0}^2 - \sigma_{\epsilon}^2)\sigma_U^2 + \beta_0^2\sigma_U^2\{\bomega\trans E(X_i\Z_i) - \wh\bomega\trans\wh\bSig_{21}\} \\
&+&  \wt\sigma_{\epsilon, H_0}^2 (1 - \wh\bomega\trans\wh\bSig_{21})
 -  \sigma_{\epsilon}^2 \{1 - \bomega\trans E(X_i\Z_i)\}.
\ese
Let  $D_1 = (\wt\sigma_{\epsilon, H_0}^2 - \sigma_{\epsilon}^2)\sigma_U^2$, $D_2 = \beta_0^2\sigma_U^2\{\bomega\trans E(X_i\Z_i) - \wh\bomega\trans\wh\bSig_{21}\}$
and $D_3 = \wt\sigma_{\epsilon, H_0}^2 (1 - \wh\bomega\trans\wh\bSig_{21})
 -  \sigma_{\epsilon}^2 \{1 - \bomega\trans E(X_i\Z_i)\}$.
For term $D_1$,
recall that
\bse
\wt\sigma_{\epsilon, H_0}^2 &=& n^{-1}\sum_{i=1}^n(Y_i - \beta^* W_i)^2 - n^{-1}\sum_{i=1}^n(\wt\bg\trans\Z_i)^2 - \beta^{*2}\sigma_U^2, \\
\sigma_{\epsilon}^2 &=& E\{(Y_i - \beta_0W_i)^2\} - E\{(\bg_0\trans\Z_i)^2\} - \beta_0^2\sigma_U^2.
\ese
Then we have
\bse
&&\wt\sigma_{\epsilon, H_0}^2 - \sigma_{\epsilon}^2 \\
&=&n^{-1}\sum_{i=1}^n(Y_i - \beta^* W_i)^2 - E\{(Y_i - \beta_0W_i)^2\} - \left[n^{-1}\sum_{i=1}^n(\wt\bg\trans\Z_i)^2 - E\{(\bg_0\trans\Z_i)^2\}\right]\\
&=&n^{-1}\sum_{i=1}^n(Y_i - \beta_0 W_i)^2 - E\{(Y_i - \beta_0W_i)^2\} - \left[n^{-1}\sum_{i=1}^n(\wt\bg\trans\Z_i)^2 - E\{(\bg_0\trans\Z_i)^2\}\right].
\ese
First, by triangle inequality and Assumption \ref{ass:model}, we know
\bse
\|\bg_0\trans\Z_i\|_{\psi_2} \leq \sum_{j=1}^{p-1}|\gamma_{0j}|\|\Z_{ij}\|_{\psi_2} \leq K\|\bg_0\|_1 \leq s_0KK_0.
\ese
Then we have
\bse
\|s_0^{-2}(Y_i - \beta_0W_i)^2\|_{\psi_1} &=& \|s_0^{-2}(\bg_0\trans\Z_i + \epsilon_i  - \beta_0U_i)^2\|_{\psi_1} \\
&\leq& 2\|s_0^{-1}(\bg_0\trans\Z_i + \epsilon_i  - \beta_0U_i)\|_{\psi_2}^2 \\
&\leq& 2\left(\|s_0^{-1}\bg_0\Z_i\|_{\psi_2} + \|s_0^{-1}\epsilon_i\|_{\psi_2} + \|s_0^{-1}\beta_0U_i\|_{\psi_2}\right)^2 \\
&\leq& 2(K_0K + s_0^{-1}K_{\epsilon} + s_0^{-1}|\beta_0|K_U)^2 \\
&\leq& K_8,
\ese
where $K_8$ is a finite constant.
Then by Bernstein inequality, for any $t>0$ we have
\bse
\Pr\left(s_0^{-2}\left|n^{-1}\sum_{i=1}^n (Y_i - \beta_0W_i)^2 - E\{(Y_i - \beta_0W_i)^2\}\right|\geq t \right) \leq 2\exp\left\{-C^{\prime\prime}\min \left(\frac{t^2}{4K_8^2}, \frac{t}{2K_8} \right)n  \right\}.
\ese
Let $t = \sqrt{n^{-1}\log p}$. Then for $n$ large enough, we have
\bse
\Pr\left(\left|n^{-1}\sum_{i=1}^n (Y_i - \beta_0W_i)^2 - E\{(Y_i - \beta_0W_i)^2\}\right|\leq s_0^2\sqrt{n^{-1}\log p}\right) \geq 1 - 2\exp\left(\frac{-C^{\prime\prime}\log p}{4K_8^2}\right).
\ese
Thus,
\be
\label{eq:D11}
\left|n^{-1}\sum_{i=1}^n(Y_i - \beta_0 W_i)^2 - E\{(Y_i - \beta_0W_i)^2\} \right| \leq s_0^2\sqrt{n^{-1}\log p}
\ee
with probability tending to $1$.
Note that under the condition given in Lemma \ref{lemma:consistency}, $s_0^2\sqrt{n^{-1}\log p} = o(1)$.
For term $n^{-1}\sum_{i=1}^n(\wt\bg\trans\Z_i)^2 - E\{(\bg_0\trans\Z_i)^2\}$, we first have
\bse
&& \left| n^{-1}\sum_{i=1}^n(\wt\bg\trans\Z_i)^2 - E\{(\bg_0\trans\Z_i)^2\} \right|  \\
&\leq& \left|n^{-1}\sum_{i=1}^n(\wt\bg\trans\Z_i)^2  - n^{-1}\sum_{i=1}^n(\bg_0\trans\Z_i)^2\right| + \left|  n^{-1}\sum_{i=1}^n(\bg_0\trans\Z_i)^2 -  E\{(\bg_0\trans\Z_i)^2\}\right| \\
&\leq& \|\wt\bg - \bg_0\|_{1}\left\|n^{-1} \sum_{i=1}^n (\wt\bg + \bg_0)\trans \Z_i\Z_i \right\|_{\infty} +  \left|  n^{-1}\sum_{i=1}^n(\bg_0\trans\Z_i)^2 -  E\{(\bg_0\trans\Z_i)^2\}\right|.
\ese
By triangle inequality, Lemma G.4 in supplementary materials and Lemma \ref{lemma:theta}, we have
\bse
\|(\wt\bg + \bg_0)\trans\Z_iZ_{ik}\|_{\psi_1} &=& \left\|\sum_{j=1}^{p-1}(\wt\gamma_j  + \gamma_{0j})Z_{ij}Z_{ik}\right\|_{\psi_1} \\
&\leq& \sum_{j=1}^{p-1}|\wt\gamma_j + \gamma_{0j}| \|Z_{ij}Z_{ik}\|_{\psi_1} \\
&\leq& 2K^2 \|\wt\bg + \bg_0\|_1 \\
&\leq&2K^2 s_0(\|\wt\bg - \bg_0\|_\infty+2\|\bg_0\|_{\infty}) \\
&\leq& 2K^2s_0(C_{\infty}C_{\lambda}s_0\sqrt{n^{-1}\log p} + 2K_0) \\
& \leq& 2K^2s_0K_0^{\prime},
\ese
with probability tending to 1,
where $K_0^{\prime}$ is a constant. The
third last inequality holds because the support of CoCoLasso estimate $\wt\btheta$ is a subset of the true support with probability going to 1.
The second last inequality used result (b) in Lemma \ref{lemma:theta_infty} and that $\|\btheta\|_{\infty}$ is bounded.
Then $\|s_0^{-1}(\wt\bg + \bg_0)\trans\Z_iZ_{ik}\|_{\psi_1} \leq 2K^2K_0^{\prime} < \infty$, for $k = 1, \dots, p-1$.
By the definition of sub-exponential norm, we know that $|E\{s_0^{-1}(\wt\bg + \bg_0)\trans\Z_iZ_{ik}\}|$ is also finite.
By Bernstein inequality and union bound inequality, for any $t>0$ we have
\bse
&&\Pr\left( \left\| n^{-1}\sum_{i=1}^n s_0^{-1}(\wt\bg + \bg_0)\trans\Z_i\Z_i - E\{s_0^{-1}(\wt\bg + \bg_0)\trans\Z_i\Z_i\} \right\|_{\infty} \geq t \right)\\
&\leq& 2p\exp\left\{ -C^{\prime\prime} \min\left(\frac{t^2}{16K^4K_0^{\prime2}}, \frac{t}{4K^2K_0^{\prime}} \right)n \right\}.
\ese
Let $t = C\sqrt{n^{-1}\log p}$. Then for $n$ large enough, we have
\bse
&&\Pr\left( \left\| n^{-1}\sum_{i=1}^n s_0^{-1}(\wt\bg + \bg_0)\trans\Z_i\Z_i - E\{s_0^{-1}(\wt\bg + \bg_0)\trans\Z_i\Z_i\} \right\|_{\infty} \leq C\sqrt{n^{-1}\log p} \right)\\
&\geq& 1- 2p\exp\left(  \frac{-C^{\prime\prime}C^2\log p}{16K^4K_0^{\prime2}}\right).
\ese
When $C^{\prime\prime}C^2/(16K^4K_0^{\prime}) > 1$, we have
\bse
\left \| n^{-1}\sum_{i=1}^n s_0^{-1}(\wt\bg + \bg_0)\trans\Z_i\Z_i\right \|_{\infty} \leq  \|E\{s_0^{-1}(\wt\bg + \bg_0)\trans\Z_i\Z_i\} \|_{\infty} + C\sqrt{n^{-1}\log p}
\ese
with probability tending to $1$.
Hence, we obtain that
\bse
\|\wt\bg - \bg_0\|_{1}\left\|n^{-1} \sum_{i=1}^n (\wt\bg + \bg_0)\trans \Z_i\Z_i \right\|_{\infty}
 &\leq& s_0\|\wt\btheta - \btheta_0\|_1\left \{\|E\{s_0^{-1}(\wt\bg + \bg_0)\trans\Z_i\Z_i\}\|_{\infty} + C\sqrt{n^{-1}\log p}\right\} \\
 &\leq& 16s_0^2\lambda\kappa^{-1}\{\|E\{s_0^{-1}(\wt\bg + \bg_0)\trans\Z_i\Z_i\}\|_{\infty}+C\sqrt{n^{-1}\log p}\} \\
 &\leq& C_1s_0^3\sqrt{n^{-1}\log p},
\ese
for some constant $C_1$ with probability tending to $1$.

For term $\left|  n^{-1}\sum_{i=1}^n(\bg_0\trans\Z_i)^2 -  E\{(\bg_0\trans\Z_i)^2\}\right|$, since $\|s_0^{-1}\bg_0\trans\Z_i\|_{\psi_2} \leq KK_0$,
then $\|s_0^{-2}(\bg_0\trans\Z_i)^2\|_{\psi_1} \leq 2 K^2K_0^2<\infty$ by Lemma G.4.
By Bernstein inequality, for any $t >0$, we have
\bse
\Pr\left( \left| n^{-1}\sum_{i=1}^n s_0^{-2}(\bg_0\trans\Z_i)^2 - E\{s_0^{-2}(\bg_0\trans\Z_i)^2\}\right| \geq t \right)
\leq 2\exp\left\{-C^{\prime\prime}\min\left(\frac{t^2}{16K^4K_0^4}, \frac{t}{4K^2K_0^2} \right)n  \right\}.
\ese
Let $t = \sqrt{n^{-1}\log p}$, then for $n$ large enough we have
\bse
\Pr\left( \left| n^{-1}\sum_{i=1}^n (\bg_0\trans\Z_i)^2 - E\{(\bg_0\trans\Z_i)^2\}\right| \leq s_0^2\sqrt{n^{-1}\log p} \right)
\geq 1 - 2\exp\left(\frac{-C^{\prime\prime}\log p}{16K^4K_0^4 } \right).
\ese
Hence, we obtain that
\be\label{eq:D12}
\left|n^{-1}\sum_{i=1}^n(\wt\bg\trans\Z_i)^2 - E\{(\bg_0\trans\Z_i)^2\}\right| \leq C_1s_0^3\sqrt{n^{-1}\log p} + s_0^2\sqrt{n^{-1}\log p}
\ee
with probability tending to $1$.
Therefore, from (\ref{eq:D11}) and  (\ref{eq:D12}),  we obtain
\be
\label{eq:D1}
|D_1| &=&  |\wt\sigma_{\epsilon, H_0}^2 - \sigma_{\epsilon}^2|\sigma_U^2\n \\
&\leq& \left(2s_0^2\sqrt{n^{-1}\log p} + C_1s_0^3\sqrt{n^{-1}\log p} \right)\sigma_U^2 \n\\
&\leq& C_2s_0^3\sqrt{n^{-1}\log p}\sigma_U^2,
\ee
with probability tending to $1$.

For term $D2$, by triangle inequality, we first have
\bse
|D_2|
&\leq& \beta_0^2\sigma_U^2 \{|\bomega\trans\{E(X_i\Z_i) - \wh\bSig_{21}\}| + |(\wh\bomega - \bomega)\trans\wh\bSig_{21}| \}\\
&\leq& \beta_0^2\sigma_U^2\{\|\bomega\|_1\|\wh\bSig_{21} - E(X_i\Z_i)\|_{\infty} + \|\wh\bomega - \bomega\|_1\|\wh\bSig_{21}\|_{\infty} \}.
\ese
In the proof of Lemma \ref{lemma:theta}, we have showed that
$\{\wh\bSig_{21} - E(X_i\Z_i)\}_j$ is sub-exponential and
$\|\{\wh\bSig_{21} - E(X_i\Z_i)\}_j\|_{\psi_1} \leq 4K^2$
 for $j=1, \dots, p-1$.
Then by Bernstein inequality, for any $t>0$ we have
\bse
\Pr\left(\|\wh\bSig_{21} - E(X_i\Z_i)\|_{\infty} \geq t\right) \leq 2(p-1)\exp\left\{-C^{\prime\prime}\min\left(\frac{t^2}{16K^4}, \frac{t}{4K^2} \right)n \right\}.
\ese
Let $t = M\sqrt{n^{-1}\log p}$, where $M>0$. Then for any $\varepsilon > 0$, there exists $M>\sqrt{16K^4/C^{\prime\prime}}$, such that
\be\label{eq:D21}
\Pr\left(\|\wh\bSig_{21} - E(X_i\Z_i)\|_{\infty} \geq M\sqrt{n^{-1}\log p}\right) \leq \varepsilon,
\ee
for $n$ large enough. Hence, $\|\wh\bSig_{21} - E(X_i\Z_i)\|_{\infty} = O_P(\sqrt{n^{-1}\log p})$.
By Assumption \ref{ass:model}, then we have
\bse
\|\bomega\|_1\|\wh\bSig_{21} - E(X_i\Z_i)\|_{\infty} \leq K_{\bomega}M\sqrt{n^{-1}\log p} \leq C_3 \sqrt{n^{-1}\log p}
\ese
for some constant $C_3$, with probability tending to $1$.
Here, $K_{\bomega}$ is a positive constant satisfying
$\|E(X_i\Z_i\trans)\{E(\Z_i^{\otimes2})\}^{-1}\|_1 \leq 
K_{\bomega}$.
Since the data is standardized, $|E(X_iZ_{ij})| = |{\rm cor}(X_i, Z_{ij})| \leq 1$ for $j = 1, \dots, p-1$. Then $\|E(X_i\Z_i)\|_{\infty} \leq 1$.
By (\ref{eq:D21}),  $\|\wh\bSig_{21}\|_{\infty} \leq
\|E(X_i\Z_i)\|_{\infty} + M\sqrt{n^{-1}\log p} \leq 1 +
M\sqrt{n^{-1}\log p}$ for some constant $M$ and $n$ large enough, with
probability tending to $1$.
By Lemma \ref{lemma:consistency}, we have
\be
\label{eq:D22}
\|\wh\bomega - \bomega\|_1\|\wh\bSig_{21}\|_{\infty} \leq \|\wh\bomega - \bomega\|_1 + \|\wh\bomega - \bomega\|_1M\sqrt{n^{-1}\log p}
\leq C_4 s^{\prime}\sqrt{n^{-1}\log p}
\ee
for some constant $C_4$, with probability tending to $1$.
Then  we obtain
\be
\label{eq:D2}
|D_2| \leq
\beta_0^2\sigma_U^2(C_3+C_4)s^{\prime}\sqrt{n^{-1}\log p}.
\ee
with probability tending to $1$.

For term $D3$, by triangle inequality, (\ref{eq:D1}) and (\ref{eq:D2}), we  have
\bse
|D_3 |
&\leq& \wt\sigma_{\epsilon, H_0}^2 |\bomega\trans E(X_i\Z_i) - \wh\bomega\trans\wh\bSig_{21}|
+ |1 - \bomega\trans E(X_i\Z_i)|\cdot|\wt\sigma_{\epsilon, H_0}^2 - \sigma_{\epsilon}^2|\\
&\leq& \wt\sigma_{\epsilon,H_0}^2(C_3+C_4)s^{\prime}\sqrt{n^{-1}\log p} + |1 - \bomega\trans E(X_i\Z_i)| C_2s_0^3\sqrt{n^{-1}\log p} \\
&\leq& (\sigma_{\epsilon}^2 + C_2s_0^3\sqrt{n^{-1}\log p})(C_3+C_4)s^{\prime}\sqrt{n^{-1}\log p}  +  |1 - \bomega\trans E(X_i\Z_i)| C_2s_0^3\sqrt{n^{-1}\log p} \\
&\leq& C_5(s_0^3 + s^{\prime})\sqrt{n^{-1}\log p}
\ese
with probability tending to $1$, where $C_5$ is a constant.
Note that $ |1 - \bomega\trans E(X_i\Z_i)|$ is  bounded,
because $\|\bomega\|_1 \leq K_{\bomega}$ and $\|E(X_i\Z_i)\|_{\infty} \leq 1$.
Therefore,
\bse
\left| \wh\sigma_{\beta\mid\bg, H_0}^2 - \sigma_{\beta\mid\bg, 0}^2 \right|
\leq |D_1| + |D_2| + |D_3| =
O_p\{(s_0^3+s^{\prime})\sqrt{n^{-1}\log p}\}.
\ese
Since we assume that the true parameter $\sigma_{\beta\mid\bg, 0}^2$ is bounded away from $0$,
$s_0^3\sqrt{n^{-1}\log p} = o(1)$ and $s^{\prime}\sqrt{n^{-1}\log p} = o(1)$, then
\bse
\left| \frac{\sigma_{\beta\mid\bg, 0}}{\wh\sigma_{\beta\mid\bg, H_0}}-1 \right| &=& \frac{1}{\wh\sigma_{\beta\mid\bg, H_0}(\wh\sigma_{\beta\mid\bg, H_0} + \sigma_{\beta\mid\bg, 0})}|\sigma_{\beta\mid\bg, 0}^2 - \wh\sigma_{\beta\mid\bg, H_0}^2|  \\
&\leq& \wh\sigma_{\beta\mid \bg, H_0}^{-2}|\sigma_{\beta\mid\bg,0}^2 - \wh\sigma_{\beta\mid\bg, H_0}^2|\\
&\leq& C_6|\sigma_{\beta\mid\bg,0}^2 - \wh\sigma_{\beta\mid\bg, H_0}^2|,
\ese
for some constant $C_6$ with probability tending to $1$.
Hence, $| \sigma_{\beta\mid\bg, 0}/\wh\sigma_{\beta\mid\bg, H_0} - 1 | =
o_p(1)$ and $\wh T_n \to N (0, 1)$ in distribution as $n$ goes to
infinity.
\end{proof}

\subsection{Proof of Corollary \ref{cor:scoretest_alt}}
\label{app:altpower}
\begin{proof}
Under local alternatives, we know  $E\{S(\beta_n, \bg_0)\} = 0$ and $\sqrt{n}\wh S(\beta_n, \wt\bg)(\wh\sigma^2_{\beta_n|\bg})^{-1/2} $ converges to standard normal distribution by
Corollary \ref{cor:scoretest}.
Then by Taylor expansion, we have
\bse
\wh T_n &=& \frac{\sqrt{n}\wh S(\beta_n, \wt\bg)}{\sqrt{\wh\sigma^2_{\beta_n|\bg}}}
 + \left\{\frac{\partial \wh S(\beta_{0n}, \wt\bg)}{\partial \beta_{0n}}\frac{1}{\sqrt{\wh\sigma^2_{\beta_{0n}|\bg}}} - \frac{\wh S(\beta_{0n}, \wt\bg)}{2(\wh\sigma^2_{\beta_{0n}|\bg})^{3/2}} \right\} \sqrt{n}(\beta^* - \beta_n)\\
 &=&  \frac{\sqrt{n}\wh S(\beta_n, \wt\bg)}{\sqrt{\wh\sigma^2_{\beta_n|\bg}}}  + \left[ E\left\{\frac{\partial S(\beta, \bg_0)}{\partial \beta} \bigg\arrowvert_{\beta =\beta_n}\right\}  \frac{1}{\sqrt{\sigma^2_{\beta_n|\bg,0}}}  - \frac{E\{S(\beta_n, \bg_0)\}}{2(\sigma^2_{\beta_n|\bg,0})^{3/2}}\right] \sqrt{n}(\beta^* - \beta_n)+ o_p(1)\\
 &=&  \frac{\sqrt{n}\wh S(\beta_n, \wt\bg)}{\sqrt{\wh\sigma^2_{\beta_n|\bg}}}  - h E\left\{\frac{\partial S(\beta, \bg_0)}{\partial \beta} \bigg\arrowvert_{\beta =\beta_n}\right\}  \frac{1}{\sqrt{\sigma^2_{\beta_n|\bg, 0}}} + o_p(1) \\
&\to& N\{-h(\sigma_{\beta}^2)^{-1/2},1\}
\ese
in distribution, where $\beta_{0n}$ is between $\beta^*$ and $\beta_n$, and $\sigma^2_{\beta_n|\bg,0}$ is the variance of the decorrelated score $S(\beta_n, \bg_0)$ under local alternatives.
Therefore, the power function converges to $\Pr\{|Z -
h(\sigma_{\beta}^2)^{-1/2}|\geq Z_{\alpha/2}\}$, where $Z$ is a standard
normal random variable.
\end{proof}

\section*{Appendix C: Proofs Regarding Confidence Interval}

\setcounter{subsection}{0}\renewcommand{\thesubsection}{C.\arabic{subsection}}

\subsection{Proof of Theorem \ref{th:asynormal2}}
\label{app:th:asynormal2}
\begin{proof}
To prove the asymptotic normality of the one-step estimator $\wh\beta$,
we first show that  $ \{\partial \wh\S(\beta, \wt\bg)/\partial\beta\}|_{\beta=\wt\beta}=1-\wh\bomega\trans\wh\bSig_{21}$ is consistent for $ E[\{\partial S(\beta, \bg_0)/\partial \beta\}|_{\beta = \beta_0}]= 1 - \bomega\trans E(X_i\Z_i)$.
By triangle inequality, we have the following decomposition
\bse
|1-\bomega\trans E(X_i\Z_i) - (1 - \wh\bomega\trans\wh\bSig_{21})| &=& |\wh\bomega\trans\wh\bSig_{21} - \bomega\trans E(X_i\Z_i)| \\
&\leq& |\bomega\trans\{\wh\bSig_{21} - E(X_i\Z_i)\}| + |(\wh\bomega - \bomega)\trans\wh\bSig_{21}| \\
& \leq& \|\bomega\|_1\|\wh\bSig_{21} - E(X_i\Z_i)\|_{\infty} + \|\wh\bomega - \bomega\|_1\|\wh\bSig_{21}\|_{\infty}.
\ese
By (\ref{eq:D21}), we know that
$\|\wh\bSig_{21} - E(X_i\Z_i)\|_{\infty} = O_P(\sqrt{n^{-1}\log p})$.
Since $\|\bomega\|_1\leq K_{\bomega}$,
then $ \|\bomega\|_1\|\wh\bSig_{21} - E(X_i\Z_i)\|_{\infty} =O_P(\sqrt{n^{-1}\log p})$.
By (\ref{eq:D22}), we have $\|\wh\bomega -
\bomega\|_1\|\wh\bSig_{21}\|_{\infty} = O_P( s^{\prime}
\sqrt{n^{-1}\log p})$.
Hence,
$|1-\bomega\trans E(X_i\Z_i) - (1 - \wh\bomega\trans\wh\bSig_{21})| \leq  O_P(s^{\prime} \sqrt{n^{-1}\log p}) = o_P(1)$.

Recall that $\wh\beta =   \wt\beta - \wh S(\wt\btheta)/  (1 - \wh\bomega\trans\wh\bSig_{21} )$.
By plugging in the expression of $\wh\beta$, we  have
\bse
n^{1/2}(\wh\beta - \beta_0) &=& n^{1/2} \left\{\wt\beta - \frac{\wh S(\wt\btheta)}{1 - \wh\bomega\trans\wh\bSig_{21}} - \beta_0 \right\}\\
&=& n^{1/2}\left[ \wt\beta - \beta_0 - \frac{1}{1 - \wh\bomega\trans\wh\bSig_{21}}\left\{\wh S(\beta_0, \wt\bg) + \frac{\partial \wh S(\beta, \wt\bg)}{\partial \beta}\bigg\arrowvert_{\beta=\beta_0} (\wt\beta -\beta_0)\right\}\right] \\
&=&n^{1/2} \left[\wt\beta - \beta_0 - \frac{1}{1 - \wh\bomega\trans\wh\bSig_{21}}\left\{\wh S(\beta_0, \wt\bg) +(1 -\wh\bomega\trans\wh\bSig_{21})(\wt\beta -\beta_0)\right\} \right]\\
&=& -\frac{n^{1/2}\wh S(\beta_0, \wt\bg)}{1- \wh\bomega\trans\wh\bSig_{21}}\\
&=&  -\frac{n^{1/2}S(\beta_0, \bg_0)+o_P(1)}{1- \wh\bomega\trans\wh\bSig_{21}}\\
&=&  -\frac{n^{1/2}S(\beta_0, \bg_0)}{1- \wh\bomega\trans\wh\bSig_{21}} + o_P(1)\\
&=&  -\frac{n^{1/2}S(\beta_0, \bg_0)}{1- \bomega\trans E(X_i\Z_i)}\frac{1- \bomega\trans E(X_i\Z_i)}{1- \wh\bomega\trans\wh\bSig_{21}} + o_P(1)\\
&=&  -\frac{n^{1/2}S(\beta_0, \bg_0)}{1- \bomega\trans E(X_i\Z_i)}\{1 + o_P(1)\} + o_P(1)\\
&=&  -\frac{n^{1/2}S(\beta_0, \bg_0)}{1- \bomega\trans E(X_i\Z_i)} + o_P(1).
\ese
The second equality holds because
the estimated decorrelated score $\wh S(\beta, \bg)$ is linear in $\beta$, then by expanding $\wh S(\wt\beta, \wt\bg)$ around $\beta_0$, we obtain
\bse
\wh S(\wt\beta, \wt\bg) = \wh S(\beta_0, \wt\bg) + \frac{\partial \wh S(\beta, \wt\bg)}{\partial \beta}\bigg\arrowvert_{\beta=\beta_0}(\wt\beta - \beta_0).
\ese
The fifth equality holds by Theorem \ref{th:asynormal1} .
The eighth equality holds because of the consistency of $1-\wh\bomega\trans \wh\bSig_{21}$ to $1 - \bomega\trans E(X_i\Z_i)$.

By Lemma \ref{lemma:clt},
we know $n^{1/2} S(\beta_0, \bg_0) \to N(0, \sigma_{\beta\mid\bg,0}^2)$ in distribution.
Hence,
\bse
n^{1/2}(\wh\beta - \beta_0) =   -\frac{n^{1/2}S(\beta_0, \bg_0)}{1- \bomega\trans E(X_i\Z_i)} + o_P(1) \to N(0, \sigma_{\beta}^2)
\ese
in distribution, where $\sigma_{\beta}^2 = \{1-\bomega\trans E(X_i\Z_i)\}^{-2}\sigma_{\beta\mid\bg,0}^2$.
\end{proof}

\section*{Supplementary  Material}

We provide additional technical details for the results in the main body of the paper in supplementary materials.

\bibliographystyle{agsm}

\bibliography{Hmem}

\end{document}